\documentclass[twocolumn]{aastex631}

\usepackage{graphicx}
\usepackage{dcolumn}
\usepackage{bm}
\usepackage{xcolor}
\usepackage{makecell} 
\usepackage{tabularx}
\usepackage{physics}
\usepackage{comment}
\usepackage{multirow}
\usepackage{enumitem}
\usepackage{hyperref}

\newcommand{\R}[0]{\mathbb{R}}
\newcommand{\C}[0]{\mathbb{C}}

\newcommand{\msun}{{\,\rm M_\odot}}

\newcommand{\K}{\,{\rm K}}

\newcommand{\kpc}{\,{\rm kpc}}

\renewcommand{\arraystretch}{1.8}
\newcolumntype{C}[1]{>{\centering\let\newline\\\arraybackslash\hspace{0pt}}m{#1}}

\def\aap{A\&A}

\def\apjl{ApJ}

\def\apjs{ApJS}

\begin{document}

\title{Probing Atomic Dark Matter with Stellar Streams in Milky Way-Mass Galaxies}

\author[0000-0002-4858-0396]{Lucas S. Mandacar\'u Guerra}
\affiliation{Department of Physics, Princeton University, Princeton, NJ 08544, USA}
\author[0000-0002-7968-2088]{Stephanie O'Neil}
\affiliation{Department of Physics, Princeton University, Princeton, NJ 08544, USA}
\affiliation{Department of Physics \& Astronomy, University of Pennsylvania, Philadelphia, PA 19104, USA
}
\author[0000-0002-8495-8659]{Mariangela Lisanti}
\affiliation{Department of Physics, Princeton University, Princeton, NJ 08544, USA}
\affiliation{Center for Computational Astrophysics, Flatiron Institute, 162 5th Avenue, New York, NY 10010, USA}
\author[0000-0002-7638-7454]{Sandip Roy}
\affiliation{Department of Physics, Princeton University, Princeton, NJ 08544, USA}
\affiliation{Department of Astronomy \& Astrophysics, University of California, San Diego, La Jolla, CA
92093, USA}
\author[0000-0003-3939-3297]{Robyn Sanderson}
\affiliation{Department of Physics \& Astronomy, University of Pennsylvania, Philadelphia, PA 19104, USA
}
\author[0000-0001-8746-4753]{Aritra Kundu}
\affiliation{Department of Physics \& Astronomy, University of Pennsylvania, Philadelphia, PA 19104, USA
}

\author[0000-0002-8354-7356]{Arpit Arora}
\affiliation{Department of Astronomy and DiRAC Institute, University of Washington, 3910 15th Ave NE, Seattle, WA, 98195, USA}

\author[0000-0003-2806-1414]{Lina Necib}
\affiliation{Department of Physics and Kavli Institute for Astrophysics and Space Research, Massachusetts Institute of Technology, Cambridge, MA
02139, USA}

\author[0000-0003-2497-091X]{Nora Shipp}
\affiliation{Department of Astronomy and DiRAC Institute, University of Washington, 3910 15th Ave NE, Seattle, WA, 98195, USA}

\author[0000-0002-6196-823X]{Xuejian Shen}
\affiliation{Department of Physics and Kavli Institute for Astrophysics and Space Research, Massachusetts Institute of Technology, Cambridge, MA 02139, USA}



\begin{abstract}
We present the first detailed analysis of the effects of dissipative dark matter on stellar streams. As a concrete example, we generate a cosmological hydrodynamic zoom-in simulation of a Milky Way-mass galaxy, assuming that the dark matter consists of Cold Dark Matter~(CDM) with a sub-component ($\sim6\%$) of Atomic Dark Matter~(ADM). The ADM subcomponent behaves as collisional, efficiently dissipative gas and allows for the formation of dense compact objects that enhance the central density of satellite galaxies, making them more resistant to tidal disruption.  We show that stellar streams with stellar mass $M_{\rm tot, \star} \gtrsim 10^{5.5} \ \msun$ form later and exhibit prolonged star formation throughout their evolution, as compared to their CDM counterparts. Changes to star formation history are reflected on the chemical tracks of the stellar stream stars, where the youngest have enhanced [Fe/H] and [Mg/Fe] in the presence of ADM. Furthermore, a population of low-mass satellites with high ADM mass fractions is identified at low pericenter distances, which may affect the population of streams at $M_{\rm tot, \star} \lesssim 10^{5.5} \ \msun$. The results of this study should generalize to other dark matter models that lead to inner-density enhancements in satellites, such as elastic self-interacting dark matter in the gravothermal collapse regime. Animations of our simulations are available \href{https://www.youtube.com/playlist?list=PL5IS42ggFs2iFXRQ8uSj1ZfRUGwdibxMM}{here}.
\end{abstract}

\keywords{Stellar streams, Dark matter, Hydrodynamical simulations, Galaxy structure}


\section{INTRODUCTION} \label{sec:intro}

The Lambda Cold Dark Matter~($\Lambda$CDM) model provides a comprehensive, hierarchical description of structure formation. The model describes dark matter~(DM) as cold and collisionless, interacting only through gravity. It predicts that small halos made up of Cold Dark Matter~(CDM) form in overdense regions and merge into more massive halos through tidal disruption. $\Lambda$CDM has been successful in describing large-scale structure~\citep{springel2008,Planck:2018vyg}. The next frontier is to test CDM on smaller scales, where it can leave an imprint on the evolution of galaxies and their substructures. 

Recent observations have greatly expanded the sample of dwarf galaxies. Surveys such as the Dark Energy Survey~(DES; \citealt{DES2005,Drlica-Wagner2020}) and DECam Local Volume Exploration (DELVE; \citealt{Drilica-Wagner2021,Tan2025}) uncovered many such galaxies around the Milky Way~(MW), with others such as the Satellites Around Galactic Analogs survey~(SAGA; \citealt{Geha2017,Mao2024,Asali2025}), the Exploration of Local VolumE Satellites survey~(ELVES; \citealt{Carlsten2022}), and the Merian survey~\citep{Pan2025} expanding the search to farther systems. Because dwarf-mass halos are typically DM-dominated, they can serve as excellent laboratories for testing CDM.  To date, several small-scale ``tensions'' have been noted (see e.g.,~\citet{Sales2022} for a review), where CDM predictions of dwarf properties do not align with observations.  These tensions motivate a careful study of the relevant astrophysical uncertainties, especially pertaining to the modeling of baryonic feedback and the effects of varying the DM model beyond CDM. 

In addition to the dwarf galaxies themselves, observations have also been mapping out their tidal remnants in the form of stellar  streams~\citep{Bonaca2025}. Many streams around the MW have been cataloged and studied~\citep{Mateu2023}, with data from surveys such as DES~\citep{Shipp2018}, the Southern Stellar Stream Spectroscopic Survey ($S^5$; \citealt{Li2019,Li_2022}), the Apache Point Observatory Galactic Evolution Experiment (APOGEE; \citealt{Majewski2017,Sheffield2021}), and $Gaia$~\citep{Gaia2018, Koposov2019, Ibata2021}. This is only the beginning; both current and upcoming surveys---such as the Nancy Grace Roman Space Telescope~\citep{Spergel2013}, the Vera Rubin Observatory~(\citealt{Ivezic2019}), and Euclid~\citep{Laureijs2011}---are expected to discover stellar streams at even lower surface brightness.

The tidal debris of a dwarf galaxy follows an orbit similar to its progenitor, tracing the potential of the main halo environment~\citep{Koposov2010, Sanders2014, Sanderson2017}. Stellar streams thus allow for precise measurements of the local gravitational potential, revealing details about the mass distribution of the host galaxy and its satellites~\citep{Shipp_2021}. In addition, they can serve as sensitive probes of non-visible subhalos in the host.  For example, repeated encounters with these low-mass subhalos can heat the stars in a stream~\citep{ibata2002, Nibauer2025}, and the direct interaction of one of these subhalos with the stream can create a visible gap in its density~\citep{Yoon_2011,Carlberg_2012, Carlberg2013, Erkal2016, Banik2018, Bonaca2019}. Moreover, a stream's chemical composition may reveal details about the star formation history of its progenitor because the rate at which its gas is depleted is sensitive to inner density~\citep{Zhang2013}. And streams act as probes for the process of tidal disruption itself, which can be subject to numerical modeling uncertainties in galaxy formation simulations~\citep[e.g.,][]{Bosch2018}.

This paper represents the first detailed study of the effects of an alternative DM model on the properties of stellar streams in cosmological simulations. Zoom-in cosmological simulations provide a refined framework for studying stellar streams and their progenitors. Recent studies have shown that such simulations can resolve streams from dwarf galaxy progenitors down to $M_{\rm tot,\star} \gtrsim 10^{5.5}  \ \text{M}_\odot$~\citep{Panithanpaisal_2021, Shipp_2023,Riley2025, Shipp2025,Kundu2025,Riley2026, Thoron}. These works find that the abundance of streams in simulations is consistent with observations, and many properties of their progenitors are analogous to those of intact satellites. Further studies on the properties of these simulated streams and their detectability have compared their spatial distribution with observations~\citep{Li_2022,Shipp_2023}. Discrepancies arise in orbital properties, which may be caused by halo-to-halo variance, but the results hint at undiscovered populations of streams with low-surface-brightness tidal tails.

We explore the specific case of dissipative DM, which falls within the broader framework of ``dark sector'' models~\citep{Bertone:2018krk}.  In these scenarios, the DM can consist of one or more particles that can interact with themselves or the Standard Model through a new dark force.  The case where these self interactions are elastic---known as Self-Interacting Dark Matter~(SIDM; \citealt{Carlson1992,spergel2000})---has been extensively explored in the literature~\citep{TULIN20181}.  Another possibility is the case where the self interactions are inelastic, or dissipative~\citep{Blennow2017,Essig2019,Shen2021,Shen2024}.  DM models that introduce interactions between particles can have significant effects on the inner densities of galaxies~\citep[e.g., ][]{Elbert2015, Shen2021, Gemmell2024, Silverman2025}, which can in turn affect their disruption. Thus, analyzing the chemodynamical properties of stellar streams can potentially reveal details about the underlying DM microphysics.

The particular model of dissipative DM we study is Atomic Dark Matter~(ADM; \citealt{Kaplan2009}), which is motivated by models with mirror symmetries~\citep{Chacko:2005pe,Chacko:2018vss}.  In this scenario, the DM acts like a collisional gas that cools and dissipates energy, akin to baryons in the Standard Model. If a subcomponent of the DM in a galaxy is in this form, it can collapse to form rotating dark disks~\citep{Fan2013, Ghalsasi2017} and/or dark compact objects, such as dark white dwarfs, mirror neutron stars, and black holes~\citep{shandera, Curtin2020, Hippert2021, Ryan:2022hku}. 

Recent studies have implemented this model into hydrodynamic simulations of MW analogs to evaluate its effects on structure formation~\citep{Roy:2023zar,Gemmell2024,Roy:2024bcu}. Here, we simulate a new parameter point of ADM, representative of the slow-cooling regime, in high-resolution simulations of MW-mass galaxies.  We then analyze the properties of the population of stellar streams found around the MW analogs and highlight the effects of ADM that are reflected at present day. We find that ADM increases halos' resistance to tidal disruption, delaying stream formation, and alters stream progenitors' star formation histories, delaying quenching.

This paper is organized as follows. Section~\ref{sec:ADM} reviews the ADM model and describes the simulation details. It also outlines the process of identifying and classifying all bright substructures formed throughout the simulations, creating a catalog of their streams, phase-mixed structures, and present-day satellites. Section~\ref{sec:props} provides an analysis of the surviving satellite populations and compares them to the progenitors of the stellar streams. Section~\ref{sec:streams} illustrates the properties of the stellar stream populations observed at $z=0$, and Section~\ref{sec:conclusion} concludes. Appendix~\ref{sec:main} compiles supplementary figures referenced throughout the paper.

\section{Methodology} \label{sec:ADM}

This work presents two cosmological hydrodynamic zoom-in simulations of MW-mass galaxies. The primary simulation, denoted \texttt{ADM-slow}, consists of baryons, CDM, and a sub-component of ADM. The second simulation, denoted \texttt{CDM}, serves as a control.  It is zoomed in on the same halo as the first, but only includes baryons and CDM. Section~\ref{subsec:cooling} starts by reviewing the ADM model and its implementation in \textsc{Gizmo}~\citep{Hopkins2015}. Then, Section~\ref{subsec:zoom-in} presents the details of the zoom-in simulations.  Section~\ref{subsec:id_streams} outlines the methodology for identifying satellite galaxies, as well as stellar streams and phase-mixed structures in the simulations. 

\subsection{ADM Implementation}\label{subsec:cooling}

ADM is comprised of a fermionic dark proton, $p'$, and dark electron, $e'$, with corresponding masses, $m_{p'}$ and $m_{e'}$. Both have equal and opposite charges under a dark electromagnetic force mediated by a massless dark photon, with coupling constant $\alpha'$. We assume that ADM constitutes a fraction $f'=\Omega_{\text{adm}}/\Omega_{\text{dm}}$ of the DM, and that it produces a dark Cosmic Microwave Background~(CMB) a factor $\xi\equiv T_{\text{cmb}'}/T_{\text{cmb}}<1$ colder than the regular CMB. This work assumes a minimal version of ADM with no dark nuclear physics; thus, $p'$ is a fundamental state.

Currently, cosmology places some of the strongest bounds on the ADM model. ADM can affect the matter power spectrum by introducing dark acoustic oscillations~(DAO)~\citep{cyr-racine2013}. DAOs, which are analogous to baryonic acoustic oscillations, are driven by pressure support from dark radiation at early times. Current observational limits constrain their amplitude: studies using the CMB power spectrum~\citep{Bansal:2022qbi} and the high-redshift ultraviolet luminosity function of galaxies~\citep{barron2025} restrict the subfraction of ADM to $f' \lesssim\mathcal{O}(10\%)$. 

\begin{figure}[t!]
\centering
\includegraphics[width=\columnwidth]{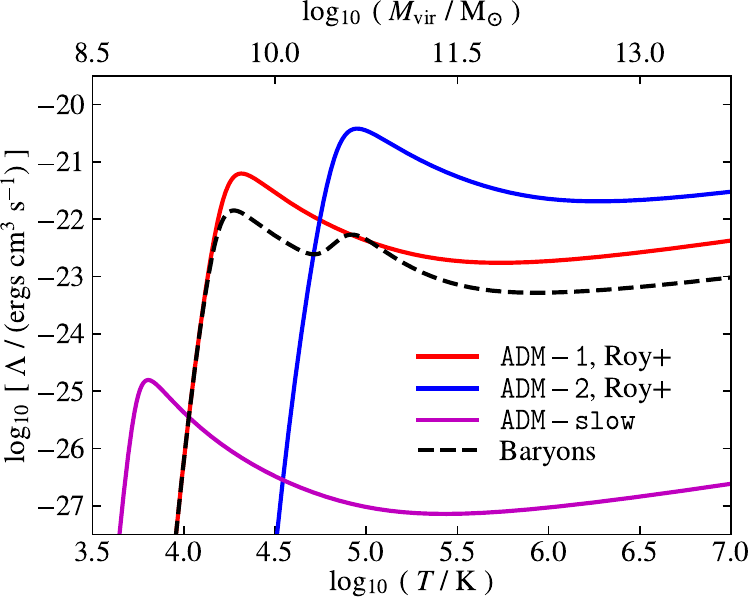}
\caption{The volumetric cooling rate $\Lambda$ versus temperature. This work simulates the \texttt{ADM-slow} model, which is shown in magenta.  Curves corresponding to the ADM parameter points simulated by \citet{Roy:2023zar} are shown in red~(\texttt{ADM-1}) and blue~(\texttt{ADM-2}). For reference, the corresponding cooling curve for baryons is shown as a dashed black line. The halo virial mass $M_{\text{vir}}$ at $z=0$ is provided on the upper axis. Compared to \texttt{ADM-1} and \texttt{ADM-2}, cooling in \texttt{ADM-slow} becomes efficient at lower temperatures, but the maximal rate is several orders-of-magnitude smaller.}
\label{fig:cooling}
\end{figure}

\begin{deluxetable*}{c|c|cccccccccc}[t!]\label{tab:params}
\tabletypesize{\small}
\renewcommand{\arraystretch}{1.2}
\tablewidth{0pt}
\tablehead{
\multicolumn{1}{c|}{Simulation} & \multicolumn{1}{c|}{Included Species} & \colhead{$\frac{\Omega_{\text{cdm}}}{\Omega_{\text{m}}}$} & \colhead{$\frac{m_{\text{cdm}}}{\text{M}_\odot}$} & \colhead{$\frac{\Omega_{\text{adm}}}{\Omega_{\text{dm}}}$} & \colhead{$\frac{m_{\text{adm}}}{\text{M}_\odot}$} & \colhead{$\frac{\Omega_{\text{b}}}{\Omega_{\text{dm}}}$} & \colhead{$\frac{m_{\text{b}}}{\text{M}_\odot}$} &
\colhead{$\frac{\alpha'}{\alpha}$} & \colhead{$\frac{m_{p'}}{m_p}$} & \colhead{$\frac{m_{e'}}{m_e}$} &
\multicolumn{1}{c}{$\frac{T_{\text{cmb'}}}{T_\text{cmb}}$} 
}
\startdata 
\texttt{CDM} & CDM+Bar. & $0.83$ & $3.49\times 10^4$ & $0$ & - & $0.17$ & $7.1\times 10^3$ & - & - & - & -\\
\texttt{ADM-slow} & CDM+ADM+Bar. & $0.78$ & $3.28\times 10^4$ & $0.06$ & $1.67\times 10^4$ & $0.17$ & $7.1\times 10^3$ & $0.22$ & $1$ & $10$ & $0.1$
\enddata\normalsize{
\caption{Summary of the simulations used in this paper. We compare simulations of the same resolution with and without ADM. All have the same total DM abundance of $\Omega_{\text{dm}}/\Omega_{\text{m}}=0.83$, but in \texttt{ADM-slow}, ADM represents $6\%$ of the DM. The symbols $m_{\text{cdm}},m_{\text{adm}}$, and $m_{\text{b}}$ refer to the particle masses for CDM, ADM, and baryons, respectively. The ADM cooling physics is defined by the dark fine-structure constant~($\alpha'$), dark proton mass~($m_{p'}$), dark electron mass~($m_{e'}$), and dark CMB temperature~($T_{\rm cmb'}$), which are shown in relation to the corresponding Standard Model values (unprimed).}\vspace{-0.8cm}}
\end{deluxetable*}
\vspace{-0.8cm}

In analogy to Standard Model hydrogen, ADM gas can cool on galactic scales, yielding distinctive phenomenology from CDM. Atomic cooling occurs through dark collisional ionization and excitation, recombination, and Bremsstrahlung. For a full discussion and summary of associated rates, see \citet{Rosenberg2017}. Figure~\ref{fig:cooling} shows the volumetric cooling curves for several ADM parameter points~(solid colored lines), comparing them to the baryonic cooling curve in dashed black. The figure also provides the halo virial mass at $z=0$ on the upper axis.\footnote{The relation between halo virial mass and temperature is redshift dependent. The threshold temperature for ADM cooling is reached at lower $M_{\text{vir}}$ at earlier times.} The ADM rates assume ionization-recombination equilibrium and no meta-galactic dark radiation background other than the dark CMB. Additionally, dark molecular cooling~\citep{Ryan:2021dis,Gurian:2021qhk,Ryan:2021tgw} is not included, as the dark molecular Jeans mass for the ADM model in \texttt{ADM-slow} is of order $\sim 10^3\msun$~\citep{Fernandez2024}, which is below our mass resolution. 

ADM cooling is largely characterized by the binding energy of dark hydrogen ($E_{\text{b}}'=\frac{1}{2}\alpha'm_{e'}c^2$, with $c$ the speed of light) and the height of the peak of the cooling rate. At temperatures below $E_{\text{b}}'$, ADM gas is essentially neutral and cooling is inefficient. For temperatures just above $E_{\text{b}}'$, the cooling rate turns on sharply; the fraction of neutral dark hydrogen is still non-zero, so collisional excitation dominates, leading to a peak in the rate. As temperatures continue to rise above this point, the rate decreases until Bremmsstrahlung takes over when $T\gg E_{\text{b}}'$, and the rate starts to slowly rise monotonically. Note that the ADM cooling curve exhibits a single peak, whereas the baryon version has two.  The second peak for baryonic cooling arises from collisional excitation of Helium, for which there is no equivalent in minimal ADM.

As described in \cite{Roy:2023zar}, ADM is implemented in \textsc{Gizmo} as a separate gas species decoupled hydrodynamically from baryons and with cooling rates given by \cite{Rosenberg2017}'s. ``Clump'' particles are formed when the local ADM gas densities in \textsc{Gizmo} are high enough for these particles to become locally self gravitating and Jeans unstable. All ADM clump particles have the same mass as ADM gas particles and only interact gravitationally. They are the sites where ADM gas will ultimately collapse to form dark compact objects at sub-resolution scales~\citep{Gurian:2022nbx}, such as dark white dwarfs~\citep{Ryan:2022hku} or sub-solar mass black holes~\citep{shandera,Fernandez2024}. 

In this work, we simulate an ADM model with parameter values $f'=6\%$, $m_{p'}=m_p$, $m_{e'}=10m_e$, $\alpha'=0.22\alpha$, $T_{\text{cmb}'} = 0.1T_{\text{cmb}}$, where unprimed values correspond to the Standard Model equivalents. These parameters are identical to those of the \texttt{slow} dwarf zoom-in simulation from \citet{Roy:2024bcu}.  Figure~\ref{fig:cooling} compares the cooling curve for this \texttt{ADM-slow} model~(magenta) to others previously studied in the literature. The \texttt{ADM-1} and \texttt{ADM-2} parameter points are representative of the aggressively cooling regime studied in~\cite{Roy:2023zar} where $\alpha'>\alpha$ and $m_{e'}<m_e$, resulting in cooling rates greater than those of the baryonic sector. For \texttt{ADM-slow}, the overall cooling rate is orders-of-magnitude smaller than that of \texttt{ADM-1} and \texttt{ADM-2}, which reduces the rate of clump formation. Additionally, because the dark hydrogen binding energy is lower, cooling is more efficient in lower-mass halos, thereby starting earlier in the simulations. 

\subsection{Simulation Procedure}\label{subsec:zoom-in}

The MW-mass simulations presented in this work are run with \textsc{Gizmo}~\citep{Hopkins2015}, which solves gravity with a Tree-PM solver~\citep{Springel2005} and hydrodynamics with a quasi-Lagrangian mesh-free finite-mass method. The target halo for the zoom-in simulations is \texttt{m12i}, selected from the standard Feedback In Realistic Environments suite~(FIRE; \citealt{Wetzel2022,Hopkins2014}). It has a halo mass of $\sim 10^{12} \ \text{M}_\odot$ and a stellar disk mass of $7.4\times 10^{10} \ \text{M}_\odot$ for a standard CDM cosmology~\citep{Hopkins2017b}. We obtain the initial transfer functions for baryons, CDM, and ADM particles using a modified version of \textsc{Class}~\citep{Blas2011,Bansal:2022qbi}. We then input these transfer functions into \textsc{Music}~\citep{Hahn2011} to generate the initial conditions at $z=100$. We assume a WMAP cosmology~\citep{Bennett2013}, with  total matter density of $\Omega_{\text{m}}=0.272$ and Hubble constant $H_0=70.2 \ \text{km} \ \text{s}^{-1} \ \text{Mpc}^{-1}$. The baryonic, CDM, and ADM abundances ($\Omega_{\text{b}}$, $\Omega_{\text{cdm}}$, and $\Omega_{\text{adm}}$,  respectively)  are summarized in Table~\ref{tab:params}.

Baryonic physics is modeled based on the prescription from the FIRE project~\citep{Hopkins2014}, specifically the ``FIRE-2'' version~\citep{Hopkins2017b}. Baryons include low-temperature cooling~\citep{Ferland1998-CLOUDY,Wiersma2009}, reionization and heating from a redshift-dependent meta-galactic ultraviolet radiation background~\citep{faucher-giguere2011,Onorbe2016} and stellar sources, star formation, as well as explicit models for stellar and supernovae feedback~\citep{Hopkins2014}. Previous studies have found that the FIRE-2 model leads to the formation of stellar stream populations with counts consistent with MW observations~\citep{Shipp_2023}.

\begin{figure*}[t!]
\centering
\includegraphics[width=\textwidth]{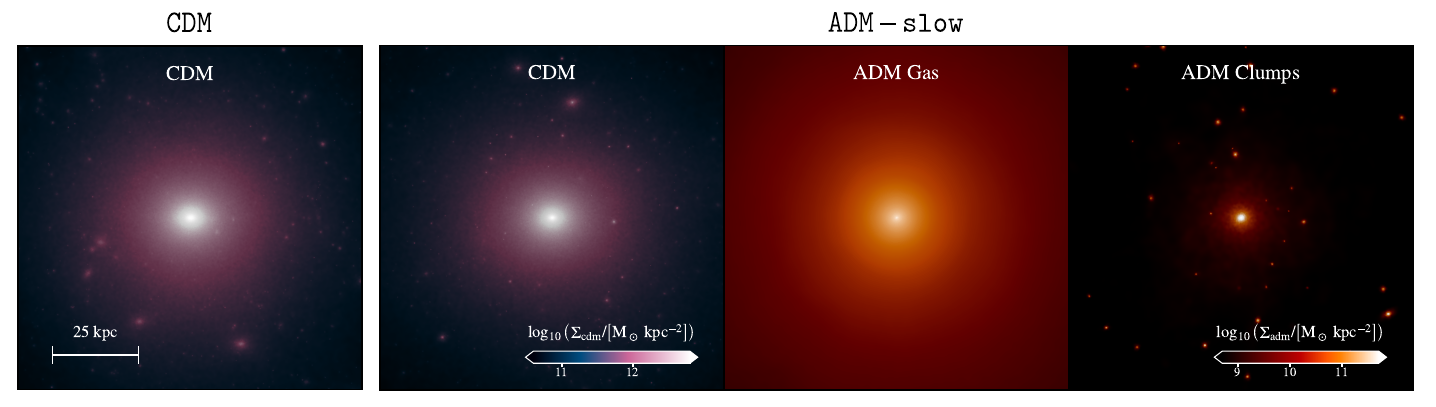}
\caption{CDM and ADM density projection plots at $z=0$, centered on the host halo. The projections for the CDM component of the \texttt{CDM} and \texttt{ADM-slow} simulations are shown by the purple colormap.  The projections for the ADM gas and ADM clump subcomponents of \texttt{ADM-slow} are shown by the red colormap. Both the ADM gas and clumps are approximately spherically distributed, with no dark disks forming. The halos in \texttt{ADM-slow} have enhanced inner DM densities.
\label{fig:main_halo}}
\end{figure*}

Table~\ref{tab:params} summarizes the parameters used to generate the \texttt{ADM-slow} and \texttt{CDM} simulations.  The number of CDM particles is the same in both simulations, but have $6\%$ lower mass in \texttt{ADM-slow} to account for the ADM subcomponent. The corresponding particle masses are $m_{\text{cdm}}=3.49\times 10^4 \ \text{M}_\odot$ and $3.28\times 10^4 \ \text{M}_\odot$ in \texttt{CDM} and \texttt{ADM-slow}, respectively. The ADM particles have mass $m_{\text{adm}}=1.67\times 10^4 \ \text{M}_\odot$, which is low enough to resolve the atomic cooling limit~\citep{Rosenberg2017,Roy:2023zar}. CDM has a force softening~\citep{Price2007} of $h_{\text{cdm}}=20 \ \text{pc}$ while ADM clumps have minimum $h_{\rm clump} = 4 \ \text{pc}$. 

To adequately resolve stellar streams, the simulations must be run at sufficiently low baryon mass and with sufficient time resolution~\citep{Panithanpaisal_2021}.  In particular, the baryon particle mass is  $m_{\rm b}=7.1\times 10^3 \ \text{M}_\odot$, and the minimum force softening for gas and stars is $h_{\text{gas}}=2 \ \text{pc}$ and $h_{\text{star}} = 4 \ \text{pc}$, respectively.\footnote{The fully conservative adaptive algorithm by~\citet{Price2007} is used to model the force softening for baryonic and ADM gas particles. Gravitational forces assume the same mass distribution as the hydrodynamic equations.} This resolution allows us to resolve streams down to stellar masses $M_{\rm tot, \star}\gtrsim 10^{5.5}\ \rm M_\odot$, placing the lower bound of
their progenitors above the present-day mass of classical
dwarf satellites of the MW.\footnote{Note that globular clusters cannot be resolved at the current simulation resolution, so they are not the progenitors of any streams studied in this work.} Each simulation stores 601 snapshots from $z=100$ to $0$, providing a time resolution of $\sim 20 \ \text{Myr}$, which corresponds to $\sim 20$ snapshots per orbit at $10 \ \text{kpc}$, allowing for dynamical analysis of tidal streams.

The left panel of Figure~\ref{fig:main_halo} shows the DM density projection for the host galaxy in the \texttt{CDM} simulation at $z=0$.  The three right-most panels show the density projections for the separate DM components of the \texttt{ADM-slow} simulation: CDM, ADM gas, and ADM clumps. Across both simulations, there are no significant changes to the virial mass and radius of the host halo or its stellar mass.  However, there are important differences in the total DM distribution, especially in the inner-most region of the host. The total DM density profiles for \texttt{CDM} and \texttt{ADM-slow} are provided in Figure~\ref{fig:main_densities} of the appendix. The presence of ADM gas and clumps in the inner region of the host halo enhances the total density within the inner $\lesssim1\ \text{kpc}$. While their concentrations have some direct contribution to this enhancement, the primary difference comes from the CDM component, which is cuspier in \texttt{ADM-slow}. This is likely caused by the presence of the ADM subcomponent.

The distribution of ADM gas forms an approximately spherical distribution with a half-mass radius of $\sim 100 \ \text{kpc}$.  The ADM gas fragments to clumps, leading to a bulge-like distribution of these dense objects concentrated within a half-mass radius of $5.2 \ \text{kpc}$ in \texttt{ADM-slow}. Clumps make up $39\%$ of the total ADM within this radius. Notably, the ADM distribution of \texttt{ADM-slow} differs from that of \texttt{ADM-1} and \texttt{ADM-2}.  In the more aggressively cooling regions of ADM parameter space, the ADM gas collapses to form a dark disk, and the clump fraction is considerably higher, closer to $95\%$~\citep{Roy:2023zar, Roy:2024bcu}. While the CDM and baryonic components in \texttt{ADM-slow} have higher resolutions than in the previous simulations, the ADM component has the same mass resolution. The differences in clump formation and distribution are thus largely attributed to the differences in cooling efficiency.\footnote{A version of \texttt{ADM-slow} with the same lower resolution as in previous studies~\citep{Roy:2023zar,Gemmell2024} is consistent with the higher-resolution simulation, with clumps making up $40\%$ of the ADM in the inner $5 \ \text{kpc}$.}

\subsection{Identifying Substructure}\label{subsec:id_streams}

To study the effects of ADM on substructure, we first need to identify satellite candidates, stellar streams, and phase-mixed structures in the simulations. Section~\ref{subsubsec:sats} begins by describing the classification of satellites. Then, Section~\ref{subsubsec:streams} discusses how to identify tidally disrupted structures with progenitors that may not be self bound at present day. The procedure relies on identifying self-bound subhalos earlier in time, assigning stars to them, and keeping track of these stars until $z=0$, when they are classified as either phase mixed or stream-like.\newpage  

\subsubsection{Satellite Galaxies}\label{subsubsec:sats}

We use \textsc{Rockstar} to identify the main DM halo and subhalos in the simulations~(\citealt{Behroozi2011}, specifically the version by \citealt{Wetzel2020}). In the \texttt{CDM} simulation, we run \textsc{Rockstar} on only the CDM particles~\citep{Samuel2020}.  In \texttt{ADM-slow}, the same process is repeated, but including the ADM clumps in addition to the CDM particles.  Halos are characterized by $R_{\text{200,m}}$, the radius that encloses $200$ times the mean matter density, and $M_{\text{200,m}}$, the total DM mass within that radius. The host is defined as the halo with the highest $M_{\text{200,m}}$ within the zoom-in region at $z=0$. Subhalos are defined as any self-bound structures identified by \textsc{Rockstar} within the virial radius ($R_{200, \text{m}}$) of the host, whether star-hosting or not.

\textsc{HaloAnalysis}~\citep{Wetzel2020halo} is used to identify star particles associated with DM subhalos, following the post-processing steps of \citet{Samuel2020}. For a given subhalo, only star particles within $\text{min}\left(0.8\times R_{\text{200,m}},30 \ \text{kpc}\right)$ of its center are considered. An additional cut is applied to the velocities of the star particles relative to the subhalo: $v\leq 2\times V_{\text{max}}$ (maximum circular velocity) and $v\leq 2\times \sigma_{\text{dm}}^{\text{3D}}$ (3D velocity dispersion of DM). Next, we keep star particles within $r<1.5\times r_{90}$ and with $v<2\times \sigma_{\star}^{\text{3D}}$, where $r_{90}$ is the radius that encloses $90\%$ of the subhalo's stellar mass and $\sigma_{\star}^{\text{3D}}$ is the 3D velocity dispersion of the subhalo's star particles. We then repeat this iteratively until the number of star particles converges at $1\%$.

Satellite candidates are defined as subhalos located within $R_{200, \text{m}}$ of the host halo at present day with at least $10$ star particles and average stellar density greater than $300 \ \text{M}_\odot \ \text{kpc}^{-3}$. This criteria follows that used by \citet{Gemmell2024}. The cuts on assigned stars and stellar density are stringent enough to ensure stellar properties remain stable across time and to exclude most unresolved stellar streams~\citep{Samuel2020,Kundu2025}. These systems have an effective minimum mass cut of $M_\star\gtrsim 10^{4.5}\ \text{M}_\odot$, set by the requirement on the number of star particles. Here, $M_\star$ represents the total stellar mass assigned to a satellite at $z=0$.

Artificial disruption is a numerical effect that can cause satellites to be excessively tidally stripped in regions of high background density. \citet{Bosch2018} found that the effect is greater at lower resolutions, for halos with fewer particles and larger force softening. Given the high resolution of our simulations and minimum force softening of $20 \ \text{pc}$, we expect these numerical effects to be minimized~\citep{Garrison2019}. However, extrapolating from ~\cite{Grand2021}, artificial disruption is likely still a factor in the ultra-faint regime. Thus, higher resolution simulations are needed to confirm trends in survivability of satellites at $M_\star\lesssim 10^5 \ \text{M}_\odot$. The addition of the ADM subcomponent in \texttt{ADM-slow} is not expected to increase the rate of artificial disruption, as it does not significantly change the background density outside of the inner $\sim1 \ \text{kpc}$ of the host. However, the enhanced central densities of subhalos may increase their resistance to tidal disruption, artificial or not~\citep{Errani2016}. Direct comparisons between \texttt{CDM} and \texttt{ADM-slow} should allow us to study the effects of ADM on tidal disruption since the two experience similar numerical effects, but higher resolutions are needed to compare the low-mass results to observations.

Throughout this study, to ensure the inner regions of a halo are converged, we adopt the criterion of \citet{Power2003}, optimized for FIRE-2 simulations by \citet{Hopkins2018}. We use the relaxation time 
\begin{equation}
    t_{\text{relax}}(r) = \frac{\sqrt{200}}{8}\frac{N}{\ln N}\left(\frac{\overline{\rho}(r)}{\rho_{\text{crit}}}\right)^{-1/2}t_0 \,,
\end{equation}
where $N$ is the number of CDM and ADM particles enclosed by $r$, $\overline{\rho}(r)$ is the mean density enclosed by $r$, $\rho_{\text{crit}}$ is the critical density of the universe, and $t_0$ is the age of the universe.  The radius of convergence is  $t_{\text{relax}}(r^{\text{conv}}_{\text{dm}}) = 0.06t_0$. \citet{Gemmell2024} demonstrated the convergence of ADM effects on satellite properties at scales above this radius. Just as in their study, we find that $r^{\text{conv}}_{\text{dm}}$ is usually smaller in simulations with ADM compared to CDM-only. This is due to the enhanced density of the inner regions of subhalos due to ADM clumps.

\begin{figure*}[t!]
\centering
\includegraphics[width=\textwidth]{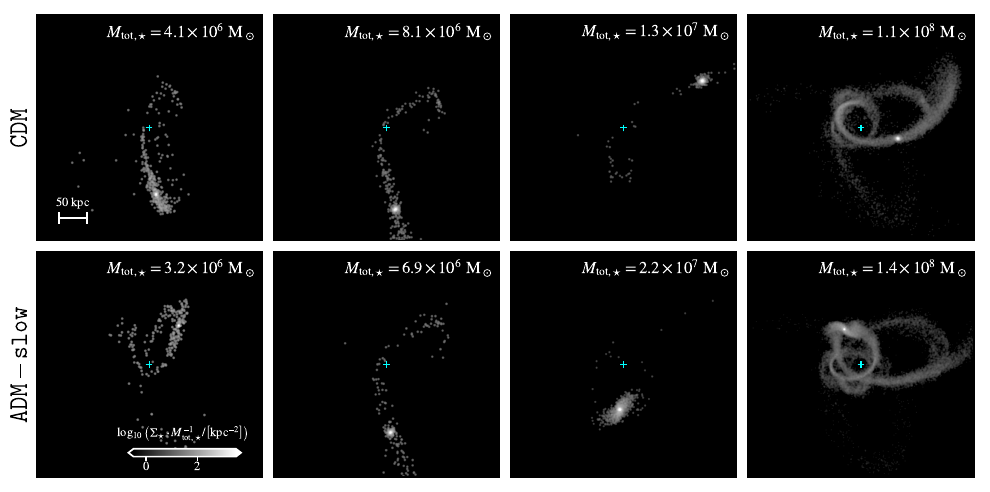}
\caption{Examples of density projections for the stars in stellar streams at $z=0$. The top row corresponds to streams in \texttt{CDM} and the bottom to \texttt{ADM-slow}. Each projection is shown from the same angle, centered at the main halo center, shown as a cyan cross. The total stellar mass of each stream at $z=0$ is shown on the corresponding plot, with mass increasing from left to right. Each column corresponds to a pair of streams across simulations with visually similar orbits and similar $M_{\rm tot, \star}$. The remaining streams are shown in Figure~\ref{fig:remaining_streams}, and examples of structures classified as phase mixed are shown in Figure~\ref{fig:pm}. Animations showing the formation of these substructures are available \href{https://www.youtube.com/playlist?list=PL5IS42ggFs2iFXRQ8uSj1ZfRUGwdibxMM}{here}.}
\label{fig:streams}
\end{figure*}

\subsubsection{Stellar Streams and Phase-Mixed Debris}\label{subsubsec:streams}

To identify stellar tidal debris in a MW-mass host, we start by determining the star particles that are stripped from a subhalo throughout its evolution. We use \textsc{consistent-trees}~\citep{Behroozi_2012} to connect subhalos across simulation snapshots. With these results, we then track each subhalo from the time it is identified by \textsc{Rockstar} to either $z=0$ or the point when it is no longer tracked by \textsc{consistent-trees} (approximating the time at which it fully merges with the host halo).\footnote{The \textsc{Rockstar} halo finder is often unable to identify halos at low pericentric radii for several snapshots, leading them to be lost by \textsc{consistent-trees} before actual tidal disruption. Alternative algorithms have been proposed to address this issue~\citep[e.g.,][]{Han2018,Diemer2024,Mansfield2024,kong2025}, which may be implemented in future studies.
} 

To ensure high fidelity in characterizing the stars that are associated with a given subhalo, we proceed as follows. Starting at some initial redshift, we select every star associated to the subhalo, as identified by \textsc{HaloAnalysis}. The initial redshift is $z_{3:1}$, the time at which the stellar mass ratio of the most massive halo to the second-most massive is $3:1$. This represents the earliest point at which the host halo becomes the dominant environment and can be distinguished from its satellites~\citep{Santistevan2020,Kundu2025}. We continue this process at each snapshot until $z=0$ or when the subhalo is lost by \textsc{consistent-trees}.  We only keep those stars that have been assigned to the subhalo for 10 consecutive snapshots or longer. This condition reduces the contamination from other nearby satellites or disk stars. To further reduce contamination, we track the evolution of the subhalo's stellar mass and remove stars that are identified in a snapshot where $M_\star$ exhibits a spurious spike. 

\begin{deluxetable*}{p{0.5\columnwidth} p{1.5\columnwidth} cc}[t!]
\renewcommand{\arraystretch}{1.2}
\tablewidth{0pt}
\tablehead{
\multicolumn{1}{c}{Substructure} & \multicolumn{1}{c}{Definition}
}
\startdata 
\centering Subhalo & Self-bound structure identified by \textsc{Rockstar} that falls within the virial radius ($R_{200, \text{m}}$) of the host by $z=0$, whether star-hosting or not. \\ 
\centering Satellite & Subhalo identified by \textsc{Rockstar} at $z=0$, with at least 10 star particles and average stellar density $>300 \ \text{M}_\odot \ \text{kpc}^{-3}$. It is labeled as a \emph{stream progenitor} if it has undergone tidal disruption and become one of the objects classified as a ``stream'' at $z=0$. It is labeled as \emph{intact} if it has not been identified as a surviving stream progenitor.\\  
\centering Phase-mixed object & Disrupted structure found within $R_{200, \text{m}}$ of the host at $z=0$, with at least 100 star particles, maximum pairwise distance $>120 \ \kpc$, and median local velocity dispersion greater than Equation~\ref{eq:criterion}. \\ 
\centering Stream & Disrupted structure found within $R_{200, \text{m}}$ of the host at $z=0$, with at least 100 star particles, maximum pairwise distance $>120 \ \kpc$, and median local velocity dispersion lower than Equation~\ref{eq:criterion}.  
\enddata
\caption{Summary of definitions used throughout this paper. Classification criteria are described in detail in Section~\ref{subsec:id_streams}. Objects with pairwise separation $<120 \ \kpc$ are assumed to be intact satellites and can be matched with structures found in Section~\ref{subsubsec:sats}.\label{tab:definitions}}\vspace{-0.5cm}
\end{deluxetable*}\vspace{-0.8cm}

We track all star particles associated with a given subhalo to $z=0$, focusing on those located within the virial radius of the host at that time.  The resulting cluster of stars constitutes the present-day stellar substructure associated with the original subhalo. Any duplicate substructures are combined in post-processing based on the fraction of star particles simultaneously assigned to both. In particular, pairs of candidates with more than $50\%$ overlap are assumed to be duplicates, while smaller fractional overlaps are inspected to ensure that they are not separate halos with a small fraction of misassigned stars.

Following \citet{Panithanpaisal_2021}, the debris from an individual subhalo is classified as a stream at $z=0$ if it has 
\begin{itemize}
\item more than $100$ associated star particles, 
\item a maximum pairwise separation between star particles greater than $120 \ \text{kpc}$, and 
\item a median stellar velocity dispersion, $\langle \sigma_{\star,\text{local}}^{\text{3D}} \rangle$, lower than a given threshold, defined below.  
\end{itemize}
The requirement on the number of star particles ensures the streams are resolved in the simulations~\citep{Kundu2025}. It effectively restricts them to $M_{\rm tot, \star} \gtrsim 10^{5.5} \ \text{M}_\odot$, where $M_{\rm tot, \star}$ represents the total mass of the stars associated with a stream at $z=0$. The distance cut between star particles minimizes the number of intact satellites that are misclassified as streams. 
We confirm that most substructures accepted by loosening this cut are intact satellites, in either \texttt{CDM} or \texttt{ADM-slow}.

The median local 3D velocity dispersion for the star particles, $\langle \sigma_{\star,\text{local}}^{\text{3D}} \rangle$, is based on their nearest neighbors in phase space. At late stages in their lifetime, streams are fully absorbed by the main halo and become phase mixed. Phase-mixed structures have higher $\langle \sigma_{\star,\text{local}}^{\text{3D}} \rangle$, providing a good criterion for separating them from streams. Specifically, we follow the stellar mass-dispersion relation computed by \citet{Panithanpaisal_2021}, which was optimized to separate these classes of objects in their CDM simulations:
\begin{equation}\label{eq:criterion}
    \langle \sigma_{\star,\text{local}}^{\text{3D}} \rangle = -5.28\log\left(\frac{M_\star}{\text{M}_\odot}\right) + 53.55 \,.
\end{equation}
Stellar substructures that fall above this line are classified as phase mixed, while those that fall below it are classified as tidal streams.  Using FIRE simulations generated at the same resolution as those in this work, \citet{Panithanpaisal_2021} identified substructures as streams or phase mixed by eye and found that Equation~\ref{eq:criterion} provided an optimal separation between the two. We visually confirm that this criterion is effective at separating streams from phase-mixed debris in \texttt{CDM} and \texttt{ADM-slow}.

We conduct a visual inspection to confirm the substructure characterization from the criteria outlined above and remove one stream candidate from each of the \texttt{CDM} and \texttt{ADM-slow} simulation. In both cases, the candidate more closely resembled an intact satellite, having fewer than $10$ isolated star particles with more than $120 \ \text{kpc}$ pairwise separation. Many of the stellar streams identified still have surviving progenitors dense enough to be accepted by our satellite candidate criteria at $z=0$. These progenitors with associated tidal tails are removed from the satellite sample, leading to our final sample of intact satellites. Note that we cannot identify which low-mass satellites may be progenitors of unresolved streams, but the cuts applied in Section~\ref{subsubsec:sats} minimize this contamination~\citep{Samuel2020,Kundu2025}. The definitions of substructure used in this study are summarized in Table~\ref{tab:definitions}.

\begin{figure*}[t!]
\centering
\includegraphics[width=0.81\textwidth]{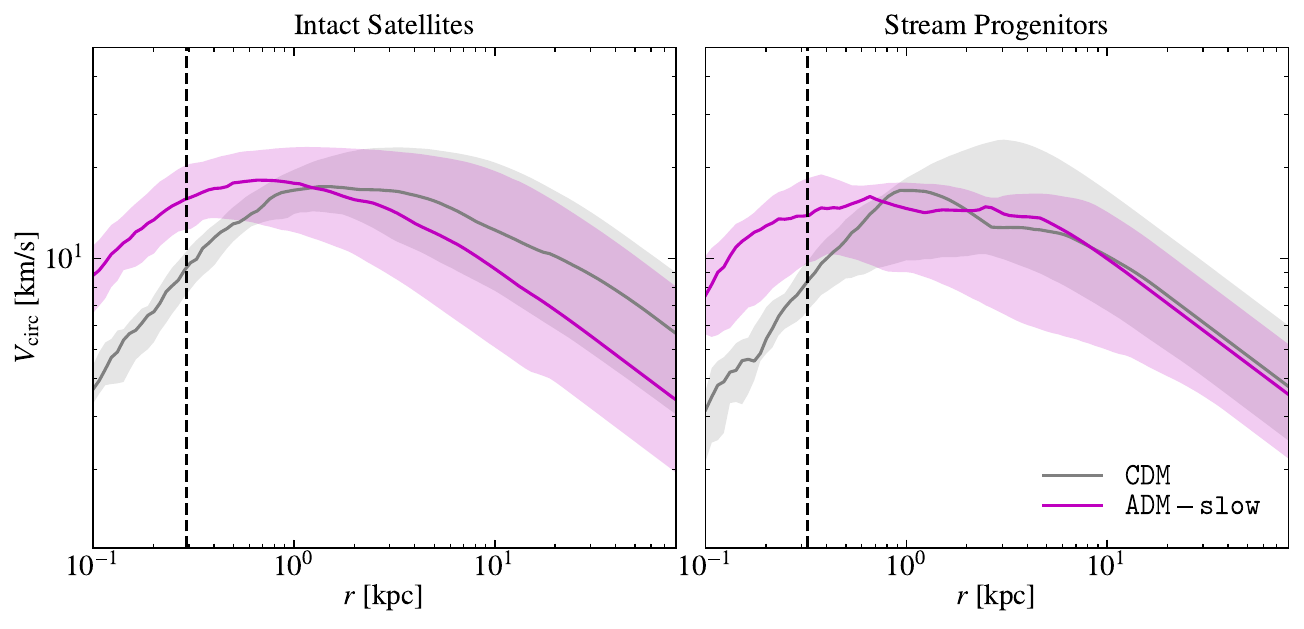} 
\caption{Median 3D circular velocity, $V_{\text{circ}}$, as a function of radius for intact satellites~(left) and stream progenitors~(right) at $z=0$. Shaded bands indicate the $68\%$ containment regions. $V_{\text{circ}}$ is calculated with all matter species in each simulation. The vertical dashed lines correspond to the maximum radius of convergence among the halos in the corresponding plot. The presence of the ADM subcomponent enhances the $V_{\text{circ}}$ in the inner regions of all halos in \texttt{ADM-slow}. This enhancement is enough to shift the maximal velocity ($V_{\text{max}}$) for both populations. On the left plot, the median in \texttt{ADM-slow} falls below the median in \texttt{CDM} at high values of $r$ due to the presence of more low-mass intact satellites in the former.}
\label{fig:Vcirc}
\end{figure*}

In the end, we identify 22 and 15 intact satellite galaxies at $z=0$ in \texttt{ADM-slow} and \texttt{CDM}, respectively. Furthermore, we find 18 phase-mixed structures in \texttt{ADM-slow} and 22 in \texttt{CDM}. Nine stellar streams are identified in both \texttt{ADM-slow} and \texttt{CDM}. Figure~\ref{fig:streams} illustrates examples of these streams. The Appendix provides the density projections for the remaining streams (Figure~\ref{fig:remaining_streams}) as well as for a subset of the phase-mixed structures (Figure~\ref{fig:pm}). Finally, Figure~\ref{fig:mass_functions} provides the present-day mass functions for all substructures in each host halo.

The stream classification scheme used in this work is sensitive to the simulation resolution. Spurious collisional heating~\citep{Ludlow2021} arises as energy equipartition is established between the DM and star particles. A consequence of this is that the CDM particles or ADM clumps in the host will artificially increase the velocity dispersion of cold stellar substructures. In \texttt{ADM-slow}, this effect should be dominated by the CDM particles, as they are more massive than the clumps. For example, we estimate (see Equation~1 of \citealt{Ludlow2021}) that a stream close to the velocity-dispersion threshold would have been characterized as phase mixed if either the CDM mass resolution were reduced to $m_{\text{cdm}}\gtrsim 10^{5} \ \text{M}_\odot$ or the ADM mass resolution were reduced to $m_{\text{adm}}\gtrsim 10^{7} \ \text{M}_\odot$ (assuming no changes in local background density). Increasing the mass resolution of our current simulation would further mitigate this effect and some of the current phase-mixed structures would likely be relabeled as streams in that case~\citep{Riley2025,Meziani}. However, all streams identified in this study should still be classified as streams even at these higher resolutions.

\section{INTACT SATELLITES AND STREAM PROGENITORS} \label{sec:props}

This section studies the effects of slow-cooling ADM on the intact satellites of MW-mass hosts at $z=0$, comparing their properties to those of the progenitor galaxies of stellar streams. Self-bound remnants of stream progenitors can still be identified by \textsc{Rockstar} at present day. Within the initial sample of satellite candidates found in Section~\ref{subsubsec:sats}, $9$ stream progenitors are identified in \texttt{ADM-slow} and $8$ in \texttt{CDM}---larger than the number of intact satellites of $M_\star\gtrsim 10^{5.5} \msun$ found in each host at $z=0$. Section~\ref{subsec:inner} reviews general properties of the intact satellite and stream progenitor density distributions in \texttt{CDM} and \texttt{ADM-slow}. This understanding informs the differences in how CDM and ADM are tidally stripped from an orbiting subhalo, as discussed in Section~\ref{subsec:stripping}. Lastly, Section~\ref{subsec:Rperi} analyzes the effects of ADM on the compactness and orbital distribution of these subhalos, aiming to inform possible effects on the stream population. 

\subsection{Subhalo Densities}\label{subsec:inner}

Although ADM is only a small fraction of the total DM in \texttt{ADM-slow}, its effects are significant in the inner-most regions of halos.  There, the density of ADM clumps deepen the gravitational potential, causing both the CDM and baryons to contract and become more centrally concentrated. The net result is that the enclosed mass near the center of a halo is typically larger in \texttt{ADM-slow} than in \texttt{CDM}. This is demonstrated in the left panel of Figure~\ref{fig:Vcirc}, which shows the median circular velocity $V_{\text{circ}}$ of the intact satellites in \texttt{CDM}~(gray) and \texttt{ADM-slow}~(magenta), as a function of distance from the satellite center. The enclosed mass within the inner $~\sim 1 \ \kpc$ is enhanced in \texttt{ADM-slow} relative to \texttt{CDM}. In this region, the CDM component comprises $\sim80\%$ of the enclosed mass, while the ADM clumps comprise $\sim20\%$. ADM gas and baryons are a negligible fraction.

The non-trivial mass fraction of ADM clumps in the central regions of the subhalos directly impacts the density profile of the CDM component. Indeed, the CDM component is more centrally concentrated in \texttt{ADM-slow} compared to \texttt{CDM} for subhalos above $M_{200, \rm m} \gtrsim 10^{8.25} \ \msun$ (see Figure~\ref{fig:sub_densities}). For less-massive subhalos, there is no significant difference in the CDM profiles; this is because these subhalos spend comparatively less time at virial temperatures above the threshold binding energy where ADM cooling becomes efficient. 

\begin{figure*}[t!]
\centering
\includegraphics[width=0.82\columnwidth]{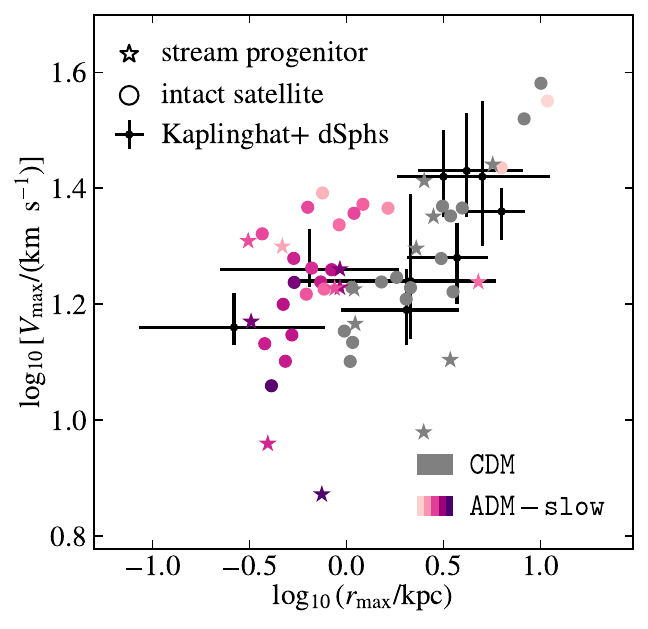} 
\includegraphics[width=\columnwidth]{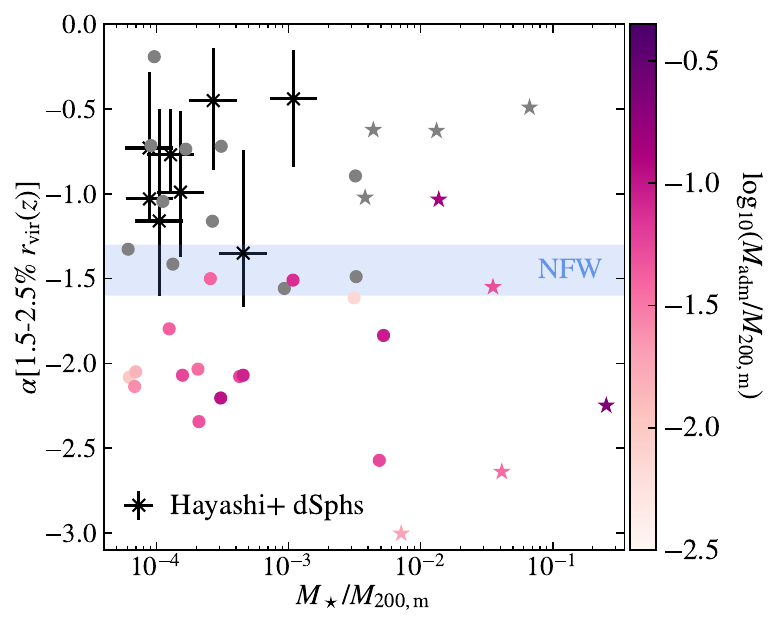}
\caption{Properties of the density profiles for the intact satellites and stream progenitors in the simulations at $z=0$. \textbf{Left}: Maximum circular velocity, $V_{\text{max}}$, versus the distance at which it is reached, $r_{\text{max}}$.  We only include subhalos with $r_{\text{max}} > r_{\text{dm}}^{\text{conv}}$. 
Intact satellites are shown as circle markers and stream progenitors are shown as star markers. The \texttt{ADM-slow} points are shaded based on the ADM clump mass fractions of the corresponding subhalos. \texttt{CDM} points are shown in gray. The black dots correspond to data from MW dwarf spheroidals, assuming an NFW DM profile~\citep{Kaplinghat2019}. The intact satellites in \texttt{ADM-slow} are shifted towards smaller $r_{\rm max}$ and slightly larger $V_{\rm max}$, compared to those in \texttt{CDM}. The stream progenitors in \texttt{ADM-slow} are also shifted towards smaller $r_{\rm max}$. \textbf{Right}: Inner DM density slope, $\alpha$ (for $\rho\propto r^\alpha$), computed over $(1.5\text{--}2.5\%)\times r_{\text{vir}}$, as a function of the stellar mass fraction, $M_\star/M_{\text{200,m}}$. The blue band shows the inner slope for NFW at $1.5\%\times r_{\text{vir}}$, based on DM-only simulations~\citep{Tollet2016}. In this panel, the black $\times$'s represent fits to data for dwarf spheroidals~\citep{Hayashi2020}. We only include satellites with $1.5\%\times r_{\text{vir}} > r_{\text{dm}}^{\text{conv}}$, with the $r_{\text{vir}}$ definition from \citet{Bryan1997}. All \texttt{CDM} points fall within or above the blue band, while almost all \texttt{ADM-slow} points fall within or below it. The ADM subcomponent thus leads to universally cuspy inner slopes.}
\label{fig:vis_slopes}
\end{figure*}

The left panel of Figure~\ref{fig:vis_slopes} shows the $V_{\text{max}}-r_{\text{max}}$ relation~(circle markers) for \texttt{CDM}~(gray) and \texttt{ADM-slow}~(colored gradient) for the intact satellites. The plot only shows systems where $r_{\rm max} > r^{\text{conv}}_{\text{dm}}$. The colored gradient corresponds to the ADM mass fraction ($M_{\rm adm}/M_{\rm 200, m}$) of each satellite. Here, $V_{\rm max}$ is the maximum circular velocity and $r_{\rm max}$ is the radius at which it is reached.  Compared to \texttt{CDM}, the intact satellites in \texttt{ADM-slow} exhibit a decrease in $r_{\text{max}}$, from $2.1^{+1.7}_{-1.0}$ to $0.6^{+0.6}_{-0.2}$, respectively. In contrast, their $V_{\rm max}$ slightly increases from $17.3^{+6.1}_{-2.6}$ to $18.2^{+5.3}_{-4.5}$. (Throughout, ranges indicate the 16-50-84$^{\rm th}$ percentiles.) These effects are consistent with Figure~\ref{fig:Vcirc}, which shows that the addition of ADM enhances the satellite $V_{\text{circ}}$ of the inner regions, often shifting the maximum to lower radii. 

The results from both simulations are roughly consistent with observations from MW dwarf spheroidals~\citep{Kaplinghat2019} which are overlaid on the left panel of Figure~\ref{fig:vis_slopes} as black points. The \texttt{ADM-slow} satellites do not extend to the highest values of $r_{\rm max}$ probed by observations, while the \texttt{CDM} satellites do not extend to the lowest values of $r_{\rm max}$.  A larger sample of MW-mass hosts is needed to understand whether this is a general trend or specific to the initial conditions used here.  Notably, the \texttt{ADM-slow} results are more reasonable than those from the aggressively cooling  ADM models studied in~\cite{Gemmell2024}, which produced satellites with significantly higher $V_{\rm max}$ than observed in data.

\begin{figure*}[t!]
\centering
\includegraphics[width=0.81\columnwidth]{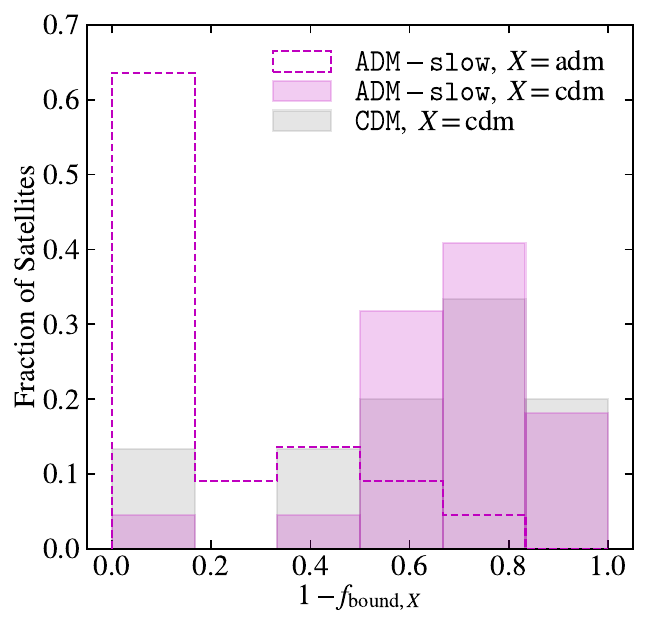}
\includegraphics[width=0.81\columnwidth]{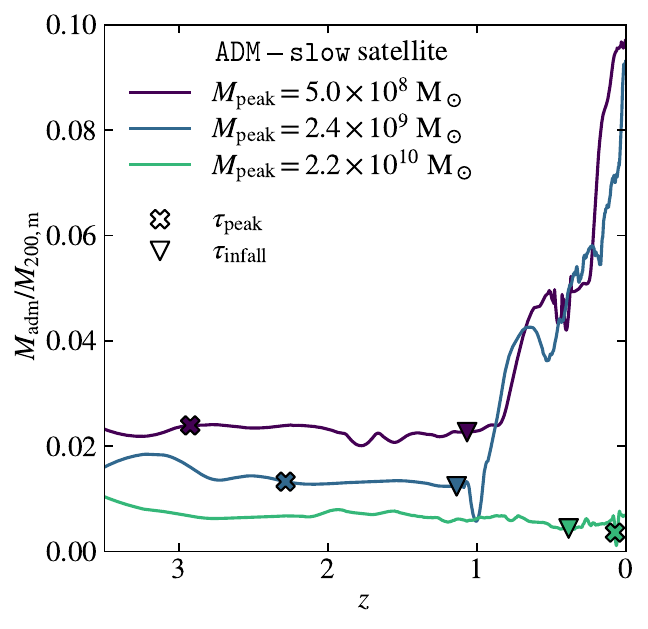}
\caption{\textbf{Left:} Distributions of $1-f_{\rm bound}$ for the CDM component and ADM clumps of intact satellites. Here, $f_{\rm bound}=M_{\rm bound}(z=0)/M_{\rm peak}$, where $M_{\rm bound}(z=0)$ is the bound mass at present day, and $M_{\rm peak}$ is the peak $M_{200,\rm m}$. Results for \texttt{ADM-slow} are shown in magenta and for \texttt{CDM}, in gray. The dashed line presents the ratios for ADM clumps. The mass-loss fraction of CDM is comparable in both simulations and significantly larger than that for ADM clumps. \textbf{Right:} ADM clump mass fraction ($M_{\text{adm}}/M_{200,\text{m}}$) as a function of redshift ($z$) for three sample intact satellites in \texttt{ADM-slow}. The curves are colored based on $M_{\text{peak}}$. The timestamps at which they reach their peak $M_\star$ ($\tau_{\text{peak}}$) are indicated as bold $\times$'s, and the timestamps at which they first fall within a virial radius ($R_{200, \rm m}$) of the host ($\tau_{\text{infall}}$) are indicated as bold triangles. The initial $M_{\text{adm}}/M_{200,\text{m}}$ of the satellites seems to be inversely correlated with $M_{\text{peak}}$, likely due to the distance from the peak in cooling rate. This fraction can significantly increase towards present time depending on the tidal stripping experienced by the satellites after $\tau_{\text{infall}}$.}
\label{fig:stripping}
\end{figure*}

Another relevant property of a subhalo's density distribution is its inner slope, $\alpha$, which is plotted in the right panel of Figure~\ref{fig:vis_slopes} versus the stellar mass ratio $M_\star/M_{200,\text{m}}$.  To obtain $\alpha$, we fit a density profile $\rho(r)\propto r^\alpha$ to the radial range $(1.5\text{--}2.5)\%\times r_{\text{vir}}$, where $r_{\text{vir}}$ is the virial radius defined by \citet{Bryan1997}. We only analyze structures with $r^{\text{conv}}_{\text{dm}}<1.5\%\times r_{\text{vir}}$. 
The blue band in Figure~\ref{fig:vis_slopes}~(right) represents the values for NFW profiles at $1.5\%\times r_{\text{vir}}$, derived from DM-only simulations~\citep{Tollet2016}. The black $\times$'s represent the best-fit results for classical dwarf spheroidals~\citep{Hayashi2020}. 

In general, the inner-slope of the satellites is significantly enhanced in \texttt{ADM-slow} compared to \texttt{CDM}, with many being cuspier than NFW.  For these two simulations, $\alpha =-2.1^{+0.5}_{-0.1}$ and $-1.1^{+0.4}_{-0.3}$, respectively.  While the enhancement to the inner slope is significant, it is less strong than in the aggressively cooling  regime~\citep{Gemmell2024}.  That being said, the simulated host in \texttt{ADM-slow} still struggles to produce the most cored satellites observed in the MW. 

The stream progenitors are indicated by the star markers in both panels of Figure~\ref{fig:vis_slopes}. Their properties are based on the output provided by \textsc{Rockstar} and \textsc{HaloAnalysis} at $z=0$. Similarly to the intact satellites, the progenitors in \texttt{ADM-slow} have lower $r_{\rm max}$ than in \texttt{CDM}. Some of these halos reach lower values of $V_{\rm max}$ than the intact satellites, likely due to significant tidal stripping. The right panel illustrates how trends in inner densities agree with the satellite populations, with significant cusps in \texttt{ADM-slow}. Both results are consistent with the right panel of Figure~\ref{fig:Vcirc}, which shows that ADM leads to an enhancement in the inner regions of the progenitors. \vspace{-0.3cm}

\subsection{Tidal Stripping of CDM vs. ADM}\label{subsec:stripping}

As demonstrated, both intact satellites and stream progenitors in \texttt{ADM-slow} can have an enhanced enclosed mass within  $\sim 1$~kpc of their centers, relative to those in \texttt{CDM}, due to the concentration of ADM clumps in this region. Here, we explore how this change in mass distribution affects the tidal stripping of the subhalos. 

The left panel of Figure~\ref{fig:stripping} provides the ADM/CDM mass fraction that is no longer bound at $z=0$ for the intact satellites in \texttt{ADM-slow} and \texttt{CDM}.  To accurately calculate the bound mass fraction, $f_{\rm bound}$, for a given species, we follow an iterative unbinding procedure~\citep[e.g.,][]{Springel2001}. We identify all particles associated with a satellite at the time it reaches its peak $M_{200,\rm m}$ and track them to $z=0$. We then identify their center with a Gaussian kernel density estimation~(KDE) and compute the kinetic energy of the particles. Next, we compute their potential with a tree method~\citep{Dehnen2000}. We then find an initial set of bound particles (total energy $<0$) and update the potential based on this set, iterating upon this process until the bound mass converges within $0.01\%$.\footnote{This procedure is carried out with the \textsc{iterative\_unbinding} method, provided at \\ \href{https://github.com/appy2806/Nbody_streams/tree/main}{https://github.com/appy2806/Nbody\_streams/tree/main}.} Comparisons between the resulting $f_{\rm bound}$ and those obtained with \textsc{Rockstar} are provided in Figure~\ref{fig:fbound}. 

In both simulations, the median CDM mass-loss fraction is $\sim69\%$.  However, the value for the ADM clumps is only $\sim 3\%$. Clearly, it is significantly harder for the clumps to be stripped than the CDM component. This is because they are more tightly bound in the central regions of the satellites.

One consequence of this is that the ADM mass fraction can increase over time, especially once the subhalo crosses the virial radius of the host. This effect is illustrated in the right panel of Figure~\ref{fig:stripping}, which displays the evolution of $M_{\text{adm}}/M_{200,\text{m}}$ as a function of redshift for three sample intact satellites in \texttt{ADM-slow}. Soon after falling within $R_{200,\text{m}}$ of the host, the two least-massive satellites experience significant tidal stripping. Since ADM clumps are more difficult to remove, $M_{\text{adm}}$ remains roughly constant, while $M_{200,\text{m}}$ decreases, so the mass fraction of ADM clumps increases until $z=0$. The most massive satellite plotted in Figure~\ref{fig:stripping}~(right) falls into the host halo environment at very late times and does not experience significant tidal stripping, thus maintaining a low $M_{\text{adm}}/M_{200,\text{m}}$. This corresponds to the satellite with highest $r_{\rm max}$ in Figure~\ref{fig:vis_slopes}.

The presence of the ADM subcomponent does not stop the stripping of the subhalos' outer layers of CDM. However, it does lead to the formation of a dense inner region, with tightly bound clumps resistant to tidal stripping. These clumps enable less-massive satellites that have undergone significant stripping to survive in \texttt{ADM-slow}, but not in \texttt{CDM}. Figure~\ref{fig:SMHMR} provides the stellar-to-halo mass relation for all the intact satellites and shows an enhancement in the number of surviving satellites with low halo mass and high $M_{\text{adm}}/M_{200,\text{m}}$ in \texttt{ADM-slow}. In particular, $5$ satellites with $M_{\rm peak}\lesssim 10^{8.5} \ \msun$ are identified in \texttt{ADM-slow}, all with $M_{\text{adm}}/M_{200,\text{m}}\sim \mathcal{O}(10\%)$, while none in this mass range are found in \texttt{CDM}.

The presence of these tightly-bound clumps might enhance the stellar population's resistance to tidal stripping. Tracing back the stream progenitors to their time of peak $M_\star$ ($\tau_{\rm peak}$), generally before tidal disruption, the range of half-mass radii for the clumps in the \texttt{ADM-slow} progenitors is $0.3^{+0.7}_{-0.2} \ \kpc$, which has significant overlap with the range of stellar half-mass radii of $0.8^{+0.6}_{-0.3} \ \kpc$. Furthermore, Figure~\ref{fig:prog_densities} shows that the deeper potential well induced by the ADM sub-component enhances the inner densities of the stellar populations of the progenitors at $\tau_{\rm peak}$. These effects can delay stream formation times, as will be discussed in Section~\ref{subsec:timescales}.

\subsection{Subhalo Orbits}
\label{subsec:Rperi}

This subsection explores the relation between subhalo compactness and the pericenters of the intact satellites and stream progenitors. The former is defined as  $R_{200,\text{m}}/R_{1/2}$, where $R_{1/2}$ is the radius that encloses $50\%$ of the total DM mass. We find the pericenters of all subhalos with \texttt{AGAMA}~\citep{Vasiliev2019}, first tracking the subhalo to identify a snapshot corresponding to a pericentric approach. We then compute the potential of the host at that snapshot. Following \citet{Arora2022}, we model the DM and hot gas ($T_{\rm gas} \geq 10^{4.5}$ K) using a multipole expansion in spherical coordinates. In this framework, angular dependence is resolved through spherical harmonics, while radial dependence is captured on a logarithmic grid. For the stellar and cold gas components ($T_{\rm gas} < 10^{4.5}$ K), we instead employ a Fourier expansion in cylindrical coordinates ($R,\phi,Z$), evaluated specifically on the meridional ($R-Z$) plane. We truncate the harmonic order for both expansion at $\ell_{\rm max} \leq 4$, achieving high fidelity in the reconstruction of halo orbits and pericentric distances~\citep{Arora2024}. For each intact satellite, we track all DM and star particles associated with the subhalo at that snapshot. For each stream progenitor, we first fit a Gaussian KDE to the positions of all the stream's star particles (as identified in Section~\ref{subsubsec:streams}) and identify the stars in the region with the highest density, corresponding to the progenitor~\citep{Shipp_2023}. The potential is then used to integrate the orbits of all identified particles, calculating their pericenters and taking the median distance to be $R_{\text{peri}}$.

\begin{figure}[t]
\centering
\includegraphics[width=\columnwidth]{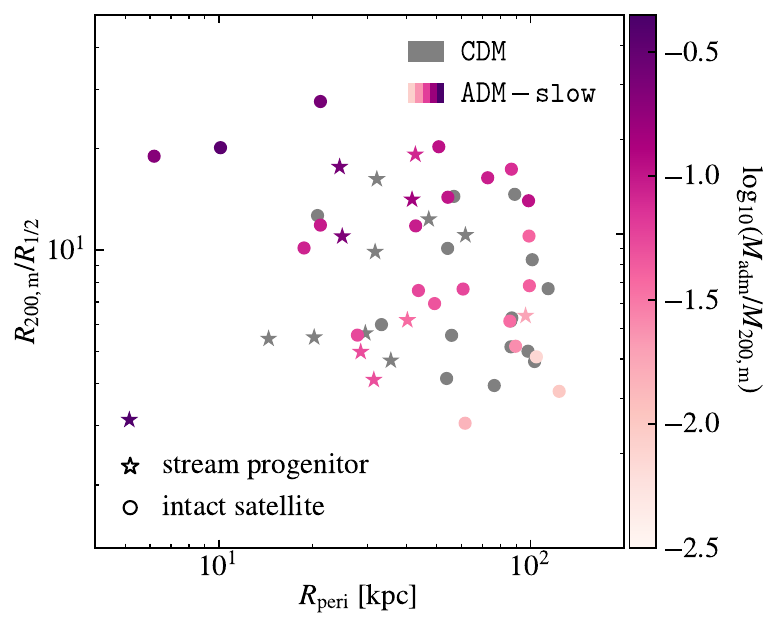}
\caption{Subhalo compactness, $R_{\text{200,m}}/R_{\text{1/2}}$, as a function of orbital pericenter distance, $R_{\text{peri}}$, at $z=0$. Intact satellites are shown as circle markers and stream progenitors are shown as star markers. The \texttt{ADM-slow} points are shaded based on the ADM clump mass fractions of the corresponding subhalos. \texttt{ADM-slow} satellites show a slight enhancement in compactness and slight decrease in average $R_{\text{peri}}$ compared to \texttt{CDM} (gray). In contrast, the progenitor distributions are comparable between the two simulations. The progenitor in \texttt{ADM-slow} with lowest $R_{\rm peri}$ has been almost entirely disrupted, decreasing its compactness.
\label{fig:vis_Rperi}}
\end{figure}

Figure~\ref{fig:vis_Rperi} shows the compactness versus $R_{\text{peri}}$ relation for the present-day satellites. There is a decrease in the median $R_{\text{peri}}$ for intact satellites in \texttt{ADM-slow} compared to \texttt{CDM}: $57.5^{+41.3}_{-36.3}$ versus $86.4^{+14.2}_{-32.5}$~kpc. Meanwhile, the satellite compactness increases in \texttt{ADM-slow} compared to \texttt{CDM}: $10.6^{+7.8}_{-5.3}$ versus $6.2^{+5.8}_{-1.5}$. The enhanced compactness is correlated with the enhanced inner density caused by the ADM clumps in the central regions of the satellites. The CDM component of the compact satellites in \texttt{ADM-slow} has undergone significant tidal stripping, as shown in Figure~\ref{fig:stripping}, but not their ADM clumps. The observed correlation between low $R_{\text{peri}}$ and high $R_{\text{200,m}}/R_{\text{1/2}}$ indicates that structures with higher compactness can survive in orbits closer to the host halo center~\citep{Errani2016}.  

The ranges of $R_{\rm peri}$ for stream progenitors are $31.4^{+11.0}_{-6.9} \ \text{kpc}$ in \texttt{ADM-slow} and $31.9^{+14.7}_{-9.3} \ \text{kpc}$ in \texttt{CDM}, which are roughly consistent. And while some progenitors in \texttt{ADM-slow} reach high compactness, many have very low values, bringing down the median. One such example is the progenitor with lowest $R_{\rm peri}$ which has been almost entirely disrupted and no longer retains a compact, spherically symmetric shape.

Note that many intact satellites reaching the lowest $R_\text{peri}$ values have $M_\star \lesssim 10^{5.5} \ \text{M}_\odot$. This is below the mass cut applied to the progenitors, so we may be missing the effect of such systems on the stream population. The fact that more of these objects are identified in \texttt{ADM-slow} may reflect changes to the population of streams at $M_{\rm tot,\star} \lesssim 10^{5.5} \ \text{M}_\odot$. This idea will be further explored in Section~\ref{subsec:orbits}.

\section{STELLAR STREAMS}\label{sec:streams}

This section studies the effects of ADM on the properties of stellar streams, analyzing all the stars identified in the procedures described in Section~\ref{subsec:id_streams}. Properties include their formation times~(Section~\ref{subsec:timescales}), metallicities~(Section~\ref{subsec:chemistry}), and orbits~(Section~\ref{subsec:orbits}). 

\subsection{Relevant Timescales}\label{subsec:timescales}

In our simulations, stellar streams are formed as satellites orbiting the MW-mass hosts are tidally disrupted. Their typical lifecycle begins with a star-forming progenitor that eventually falls into the host. At some point, either before or after infall, this progenitor is quenched. Eventually, usually near a pericentric approach, it experiences tidal compression, followed by tidal disruption. 

\begin{deluxetable}{c|cc}[t!]\label{tab:times}
\renewcommand{\arraystretch}{1.2}
\tablewidth{\columnwidth}
\tablehead{
\multicolumn{1}{c|}{} & \colhead{\hspace{0.7cm}$\Delta \tau_{\text{i,s}}$ [Gyr]}\hspace{0.7cm} & \colhead{$\Delta \tau_{\text{i,p}}$ [Gyr]\vspace{-0.25cm}}\hspace{0.7cm}\\ 
\multicolumn{1}{c|}{} & \colhead{\hspace{0.7cm}$\tau_{\text{infall}}-\tau_{\text{stream}}$}\hspace{0.7cm} & \colhead{$ \tau_{\text{infall}}-\tau_{\text{peak}}$}\hspace{0.7cm}
}
\normalsize
\startdata 
\texttt{CDM} & $1.1^{+3.8}_{-0.6}$ & $0.0^{+1.2}_{-0.8}$\\
\texttt{ADM-slow} & $2.8^{+2.6}_{-2.3}$ & $0.6^{+1.8}_{-1.5}$
\enddata
\caption{Differences in relevant timescales that dictate stream properties in our simulations: first infall time, $\tau_{\text{infall}}$, peak $M_\star$ time, $\tau_{\text{peak}}$, and stream formation time, $\tau_{\text{stream}}$, for all streams found at present day. Median and $68\%$ containment values are shown for $\Delta \tau_{\text{i,s}}\equiv \tau_{\text{infall}}-\tau_{\text{stream}}$ and $\Delta \tau_{\text{i,p}}\equiv\tau_{\text{infall}}-\tau_{\text{peak}}$. The ranges indicate that stream progenitors in \texttt{ADM-slow} reach their peak $M_\star$ and are tidally disrupted later than those in \texttt{CDM}, on average.}\vspace{-1cm}
\label{tab:timescales}
\end{deluxetable}\vspace{-0.4cm}

We keep track of the time each stream progenitor reaches its peak $M_\star$~($\tau_{\text{peak}}$), as well as two other relevant timestamps: first infall time~($\tau_{\text{infall}}$) and stream formation time~($\tau_{\text{stream}}$). $\tau_{\text{infall}}$ is when a subhalo tracked with \textsc{consistent-trees} first falls within $R_{\text{200,m}}(z)$ of the host. $\tau_{\text{stream}}$ is the time of tidal compression before tidal disruption. It is obtained by computing the moment-of-inertia tensor $I_{ij}$ for a given stream progenitor (starting at $\tau_{\rm infall}$) using coordinates centered at the assigned stars' center of mass.  We keep track of the ratio of its maximum and minimum eigenvalues at each snapshot.
Compression leads to a local minimum in the ratio of maximum and minimum eigenvalues of $I_{ij}$ ($\lambda_{\text{max}}/\lambda_{\text{min}}$), right before a global maximum during tidal disruption~\citep{Panithanpaisal_2021}. The local minimum is defined as $\tau_{\text{stream}}$. The moment-of-inertia tensor is sensitive to a few star particles located far from the center of mass. To avoid spurious peaks in $\lambda_{\text{max}}/\lambda_{\text{min}}$, we only consider particles within $2\times r_{99}$ at any snapshot, where $r_{99}$ is the radius that encloses $99\%$ of the substructure's stellar mass.

Here, we evaluate the effects of ADM on the distribution of these three timescales. Table~\ref{tab:times} provides the median and $68\%$ containment values for the relative times $\Delta \tau_{\text{i,s}}\equiv\tau_{\text{infall}}-\tau_{\text{stream}}$ and $\Delta \tau_{\text{i,p}}\equiv\tau_{\text{infall}}-\tau_{\text{peak}}$.\footnote{Note that \citet{Panithanpaisal_2021} used different criteria for assigning star particles to the stellar streams than we do. We take further steps to reduce the contamination from nearby substructure and sample from a wider time window during the progenitors' evolution. We caution the reader that it is thus difficult to directly compare the results of Table~\ref{tab:timescales} with that paper.} Compared to \texttt{CDM}, streams in  \texttt{ADM-slow} typically take longer to form after a satellite's infall. The median value of $\Delta \tau_{\text{i,s}}$ is a factor of $2.5\times$ greater for \texttt{ADM-slow}, and its 84$^{\rm th}$ percentile is also higher.  This is likely driven by the fact that the subhalos are cuspier in the presence of the ADM sub-component and thus more resistant to tidal disruption~\citep{Errani2016}. Furthermore, the concentration of ADM clumps at the center of the subhalo increases the potential well felt by the baryons, making it harder to tidally remove them.

There is also an increase in the median and $84^{\text{th}}$ percentile values of $\Delta \tau_{\text{i,p}}$ in \texttt{ADM-slow}. Since $\tau_{\text{peak}}$ reflects the progenitors' quenching time, this shift indicates that the stream progenitors in \texttt{ADM-slow} keep forming stars for longer after falling into the host, compared to \texttt{CDM}. This could be a consequence of the enhanced inner density of the \texttt{ADM-slow} halos, which could allow them to hold on to gas for longer, continuing star production. These results would be consistent with those of \citet{Zhang2013}, who found that cluster galaxies with low inner densities are more easily depleted of their gas.

\subsection{Chemical Properties}\label{subsec:chemistry}

The observation that many progenitors in \texttt{ADM-slow} exhibit prolonged star formation can potentially leave an imprint on the chemical properties of stellar streams. Throughout a galaxy's star formation history, iron is formed over long periods of time by Type~Ia supernovae. As a consequence of this, later generations of stars present higher iron abundances, [Fe/H]. This subsection demonstrates that the chemical composition of a stream at $z=0$ can serve as a proxy for the effects of ADM on its satellite progenitor. 

\begin{figure}[t!]
\centering
\includegraphics[width=0.81\columnwidth]{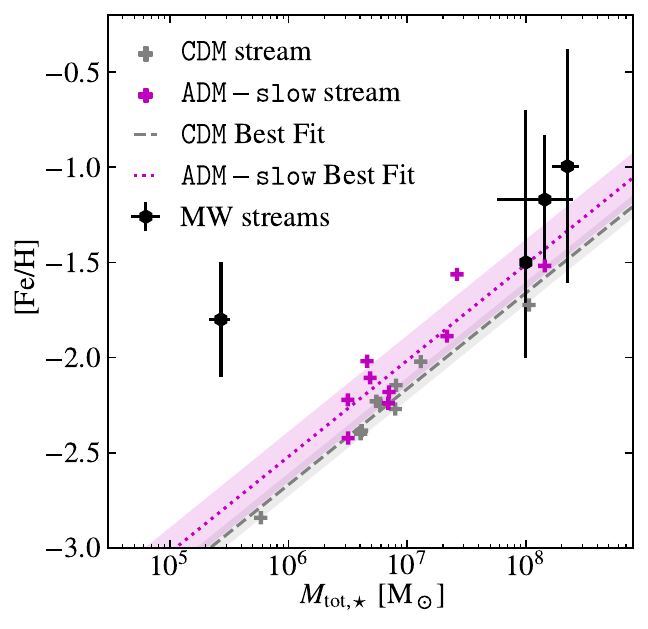}
\caption{Iron abundance, [Fe/H], versus total stellar mass, $M_{\rm tot, \star}$, for streams (crosses) at $z=0$. The \texttt{ADM-slow} points are shown in magenta and the \texttt{CDM} points in gray. The magenta and gray lines correspond to the best fit for the streams in the respective simulations, and the shaded bands correspond to the standard deviation of the residuals. The black dots correspond to MW streams: Orphan-Chenab~\citep{Mendelsohn2022,Hawkins2023}, Helmi~\citep{Koppelman2019}, Sagittarius~\citep{Mucciarelli2017,Gibbons2017,Deason2019}, and Gaia-Enceladus~\citep{Feuillet2020,Lane2023}. Note that discrepancies between  FIRE-2 modeling of chemical abundances and observations are expected~\citep{Hopkins2022}. The streams in \texttt{ADM-slow} exhibit slightly higher [Fe/H] than those in \texttt{CDM}.}
\label{fig:FeH}
\end{figure}

Figure~\ref{fig:FeH} shows the iron abundance [Fe/H] versus total stellar mass $M_{\rm tot, \star}$ for the present-day streams in \texttt{CDM} and \texttt{ADM-slow}.\footnote{The trends are similar even if plotted back to $\tau_{\rm peak}$, as almost no new stars are formed since then, and stars of all ages are lost.} We convert the iron abundance obtained from the simulations from mass to number fraction, normalizing by the solar value from \citet{Asplund2009}. The results for each stream are shown in Figure~\ref{fig:FeH} with crosses: gray for \texttt{CDM} and magenta for \texttt{ADM-slow}. 

To compare the trends in both simulations, we fit the streams from each with a linear function: 
\begin{equation}
    [\text{Fe/H}] = a\log_{10}(M_{\rm tot, \star}/\text{M}_\odot) + b \, .
\end{equation}
The shaded bands about the best-fit lines in the figure represent the standard deviation of the residuals for the fits. While the best-fit slopes are the same for \texttt{CDM} and \texttt{ADM-slow} ($a=0.50$), the intercepts are different, with $b=-5.69$ for the former and $b=-5.55$ for the latter. Thus, the streams in \texttt{ADM-slow} exhibit a small increase in iron abundance across stellar mass, compared to those in \texttt{CDM}.

\begin{figure*}[t!]
\centering
\includegraphics[width=0.85\textwidth]{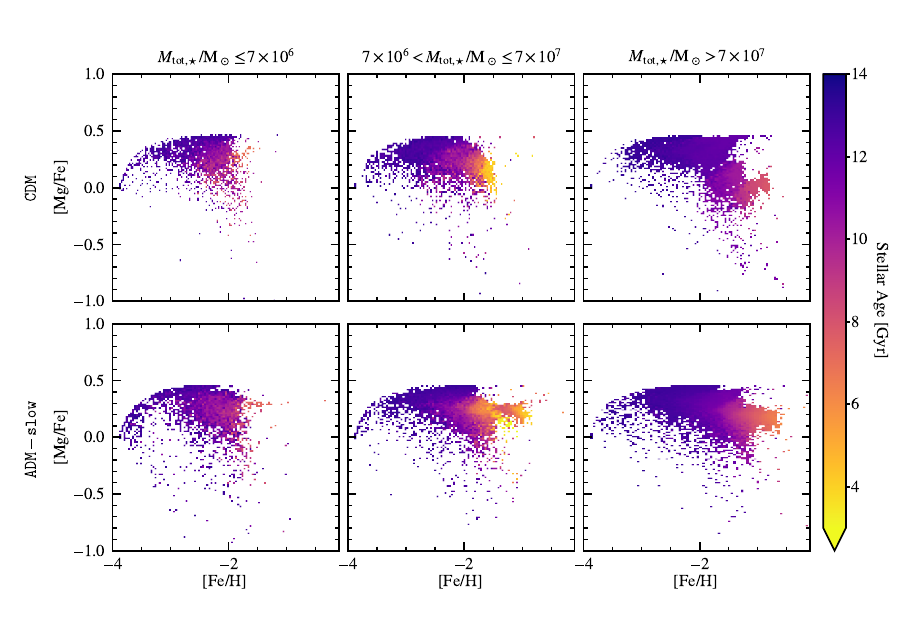}
\caption{2D projections of [Mg/Fe]$-$[Fe/H] tracks for all stars associated with streams at $z=0$. The results for \texttt{CDM} are shown in  the top row, while those for \texttt{ADM-slow} are provided in the bottom row. Each column corresponds to a different mass bin: $M_{\rm tot, \star} \leq 7\times 10^6\,\text{M}_\odot$ (left), $7\times10^6\,\text{M}_\odot < M_{\rm tot, \star} \leq 7\times 10^7\,\text{M}_\odot$ (middle), and $M_{\rm tot, \star} > 7\times 10^7\,\text{M}_\odot$ (right). The right-most column represents the most massive stream from each simulation, which has significantly more stars than all the others. The colors correspond to the mean stellar age. The \texttt{ADM-slow} streams typically contain younger stars with enhanced $\alpha$-abundance relative to those in \texttt{CDM}. See Figure~\ref{fig:MgFe} for a version of these panels plotted as superposed contours of the \texttt{CDM} and \texttt{ADM-slow} results.}
\label{fig:MgFe_ages}
\end{figure*}

For context, the black points in Figure~\ref{fig:FeH} provide observational data for four MW streams: Orphan-Chenab~\citep{Mendelsohn2022,Hawkins2023}, Helmi~\citep{Koppelman2019}, Sagittarius~\citep{Mucciarelli2017,Gibbons2017,Deason2019}, and Gaia-Enceladus~\citep{Feuillet2020,Lane2023}. 
We caution the reader that there are many sources of uncertainty in chemical modeling in simulations that make direct comparisons to observations difficult~\citep{Hopkins2018,Panithanpaisal_2021}. Indeed, discrepancies between observations and FIRE-2 modeling are expected~\citep{Hopkins2022}. The updated models of supernova rates in FIRE-3 greatly increase metal production at early times~\citep{Hopkins2022}. Since most stars associated with the streams are formed very early, these updates can potentially shift the simulation results upwards. Discrepancies in modeling of Type Ia supernovae delay time distribution lead to a vertical shift in the distribution without a change in scatter~\citep{Escala2018}. Resolution effects also add to the uncertainty, with satellites simulated at higher resolutions showing increased iron abundance~\citep{Wheeler2019}.

The small shift towards higher [Fe/H] observed for \texttt{ADM-slow} streams could reflect the fact that their progenitors typically undergo a longer period of star formation (see Section~\ref{subsec:timescales}). This corroborates the idea that the cuspier profiles in \texttt{ADM-slow} hold on to baryonic gas for longer, prolonging star formation. The $68\%$ range of $\Delta\tau_{\text{\text{i,p}}}$ is also greater in \texttt{ADM-slow}, as noted in Table~\ref{tab:times}, which could explain the increased scatter in [Fe/H]. This scatter (at a given $M_{\rm tot, \star}$) has been found to be associated with the ages of the progenitors~\citep{Riley2026}.

The $\alpha$-elements can also provide important information regarding stream formation and evolution. While Fe is created by Type~Ia supernovae over long periods of time, $\alpha$-elements such as Mg are mainly produced early in a galaxy's star formation history by Type~II supernovae. The relative [Mg/Fe] abundance of a galaxy's stars is thus high early on, but decreases throughout its formation history~\citep{Tolstoy2009}. The standard tracks in [$\alpha$/Fe] versus [Fe/H] space have been proposed as a framework to model the chemical composition of satellites, reflecting the stellar mass and accretion history~\citep{Lee2015}. This framework has been applied to intact and disrupted satellites in FIRE simulations, showing how their stellar masses and quenching times are reflected in the chemical tracks~\citep{Panithanpaisal_2021, Cunningham2022}.

Figure~\ref{fig:MgFe_ages} provides the [Mg/Fe]$-$[Fe/H] chemical tracks for \texttt{CDM}~(top row) and \texttt{ADM-slow}~(bottow row), colored based on stellar age.  From left to right, the three columns correspond to increasing stream stellar mass: $M_{\rm tot, \star} \leq 7\times 10^6 \ \text{M}_\odot, \ 7\times 10^6\ \text{M}_\odot < M_{\rm tot, \star} \leq 7\times 10^7\ \text{M}_\odot$, and $M_{\rm tot, \star} > 7\times 10^7\ \text{M}_\odot$. The high-mass bin only contains one stream for each simulation, which has significantly more stars than all the others. 

\begin{figure*}[t!]
\centering
\includegraphics[width=0.81\textwidth]{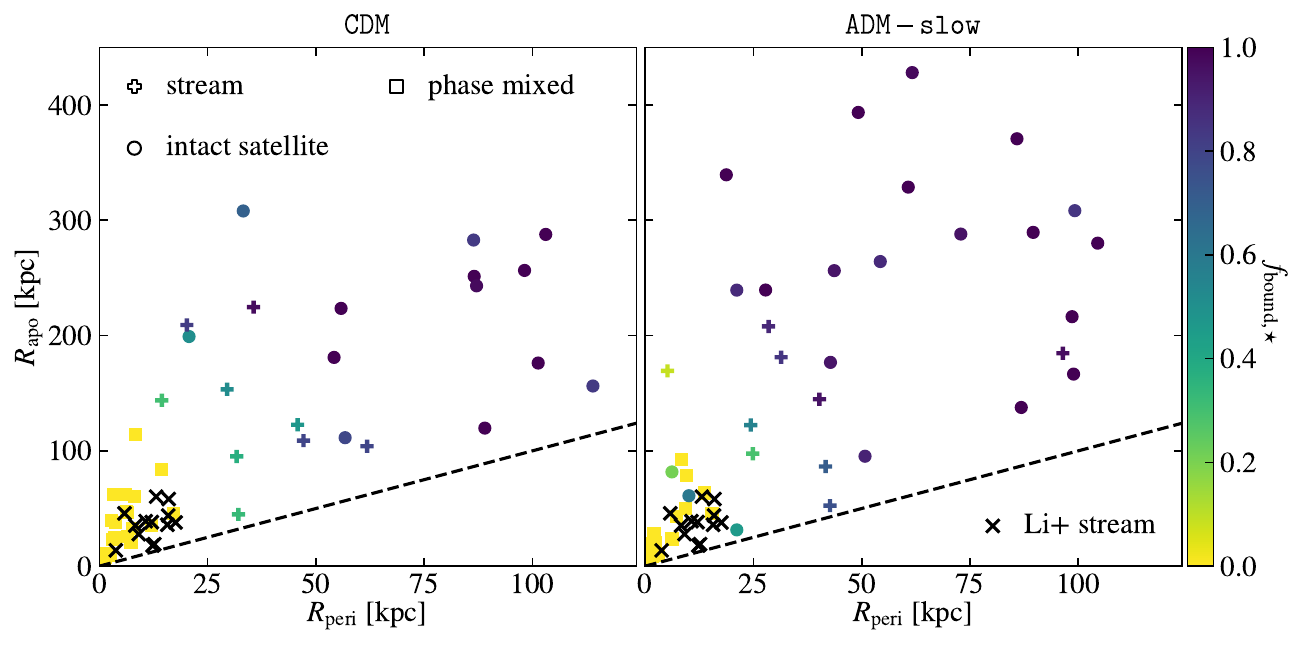}
\caption{Apocenter distance, $R_{\text{apo}}$, versus pericenter distance, $R_{\text{peri}}$, for streams (crosses), phase-mixed structures (squares), and intact satellites (circles) at $z=0$. The \texttt{CDM} results are shown on the left and \texttt{ADM-slow} on the right. Points are colored based on their bound stellar mass fraction ($f_{\rm bound,\star}$) at $z=0$. The effective stellar mass cut for intact satellites is $M_\star\gtrsim 10^{4.5}\msun$, while for disrupted systems it is $M_{\rm tot, \star}\gtrsim 10^{5.5}\msun$. The black $\times$'s correspond to MW streams from \citet{Li_2022}. In general, $f_{\rm bound,\star}$ increases moving to further orbits. A few objects identified as intact satellites lie near MW observations in \texttt{ADM-slow}, but they have low $f_{\rm bound,\star}$, indicating they may be the progenitors of unresolved streams~\citep{Riley2025}}
\label{fig:Rapo_Rperi}
\end{figure*}

For the $10\%$ youngest stars in each mass bin (which correspond to different times in each simulation), there is a universal increase in [Mg/Fe] for \texttt{ADM-slow}. For the lowest-mass streams, median [Mg/Fe] increases from $0.2$ in \texttt{CDM} to $0.26$ in \texttt{ADM-slow}. For the middle bin, the increase is from $0.17$ to $0.23$. And in the highest-mass bin, the increase is from $0.07$ to $0.18$. These trends are especially notable considering that the stellar ages in the $10^{\rm th}$ percentile are typically lower in \texttt{ADM-slow} than \texttt{CDM}, so lower [Mg/Fe] is expected of these younger stars. 

Gas outflows will be better confined in subhalos with a deepened potential well and higher escape velocity, enhancing the duty-cycle of bursty star formation and leading to the transition to steady star formation~\citep[e.g.,][]{Hopkins2023b}. ADM creates this environment in the stream progenitors in \texttt{ADM-slow}, which indeed show more frequent bursts of star formation with shorter periods of quiescence in between. (As an example, see Figure~\ref{fig:SFR}, which shows the chemical evolution and star formation rates as a function of lookback time for the most massive stream in each simulation.) To quantify this effect, we count the number of $10 \ \text{Myr}$ bins with nonzero star formation and divide by the total number of bins between the formation of the oldest and youngest stars. The median value of this ratio is $0.18$ for the progenitors in \texttt{CDM} and $0.25$ for \texttt{ADM-slow}, so the latter do indeed have shorter periods of quiescence. Bursts of star formation lead to momentary increases in [Mg/Fe], which may be reflected on the chemical tracks~\citep{Patel2022}. The frequent bursts of star formation in \texttt{ADM-slow} provide a consistent source of Type II supernovae, resulting in the increased values of [Mg/Fe] observed for its youngest stars. Long periods of quiescence may also be reflected as discontinuities in the [Mg/Fe]--[Fe/H] chemical tracks~\citep{Ting2025}, as illustrated in the high-$M_{\rm tot, \star}$ \texttt{CDM} bin of Figure~\ref{fig:MgFe_ages}. ADM may thus make these discontinuities less common.

\subsection{Orbital Properties}\label{subsec:orbits}

Lastly, we explore the orbital properties of the stellar streams in \texttt{CDM} and \texttt{ADM-slow} at present day. The procedure used to calculate the pericenter distances of intact satellites and streams is described in Section~\ref{subsec:Rperi}. To find the apocenter, we compute the potential at $z=0$ and integrate the orbits of the same associated particles. 

We also evaluate the orbits of the phase-mixed structures identified in the simulations. Because the progenitors of these structures are completely dissolved, the KDE method used for streams is not reliable here.  Therefore, we  estimate the pericenter and apocenter using every associated star particle in the phase-mixed structure.  This leads to more uncertain orbit projections: the standard deviation on $R_{\text{peri}}$ is $\sim 5 \ \text{kpc}$ for the population of stars of a given phase-mixed structure, while for streams the spread is closer to $\sim 1\, \text{kpc}$. 

Figure~\ref{fig:Rapo_Rperi} plots the apocenter, $R_{\text{apo}}$, versus pericenter, $R_{\text{peri}}$, for all substructures found in the simulations. Points are colored based on their bound stellar mass fraction, $f_{\rm bound,\star}$. Using the same procedure as in Section~\ref{subsec:stripping}, we estimate the present-day $f_{\rm bound,\star}$ for all substructures in \texttt{ADM-slow} and \texttt{CDM}. There is a strong correlation between $f_{\rm bound,\star}$ and orbit, with structures on closer orbits having lower bound fractions, consistent with the findings of ~\citet{Shipp2025}.\footnote{\citet{Shipp2025} estimate $f_{\rm bound,\star}$ with \textsc{SUBFIND}~\citep{Springel2001}, using different procedures to assign particles to halos. Furthermore, their classification criteria takes structures with $f_{\rm bound,\star}>0.97$ to be intact~\citep{Riley2025}. Following this criteria, one stream in \texttt{CDM} would be classified as intact, and several low-mass intact satellites in both simulations would be considered disrupted.} 

As described in Section~\ref{subsec:Rperi}, the stream progenitors in both simulations have similar orbital distributions. Figure~\ref{fig:Rapo_Rperi} now shows the results for all streams in the simulations, including those without an identified progenitor, and the results do not change significantly compared to Figure~\ref{fig:vis_Rperi}. In addition, the median value of $f_{\rm bound,\star}$ for streams is higher in \texttt{ADM-slow} than in \texttt{CDM} ($0.75$ compared to $0.52$), which indicates that the steeper potentials in \texttt{ADM-slow} may allow streams to keep a higher fraction of stars bound at $z=0$. \citet{Shipp2025} find that $f_{\rm bound,\star}$ is sensitive to very small changes in orbits, so higher statistics are necessary to fully characterize this effect.

As discussed in Section~\ref{subsec:Rperi}, ADM allows compact, low-mass satellites to be identified at low $R_{\text{peri}}$. This is reinforced by Figure~\ref{fig:Rapo_Rperi}, which shows three intact satellites in low-$R_{\rm peri}$, low-$R_{\rm apo}$ regions for \texttt{ADM-slow}. These satellites also have very high $M_{\rm adm}/M_{200,\rm m}$ ($\gtrsim 0.1$). However, despite being accepted by the satellite criteria in Section~\ref{subsubsec:sats} and not the streams criteria in Section~\ref{subsubsec:streams}, they have low values of $f_{\rm bound,\star}$ and may represent the progenitors of unresolved streams~\citep{Shipp2025}. Nevertheless, the fact that these self-bound structures are only identified in \texttt{ADM-slow} hints that ADM may prevent a population of progenitors from being entirely dissolved in the host halo environment, possibly keeping their associated streams coherent for longer. Higher-resolution simulations with classification criteria tuned to these smaller and lower-$M_{\rm tot, \star}$ substructures are necessary to determine the effects of ADM on the population of streams at $M_{\rm tot, \star}\lesssim 10^{5.5} \msun$.

The phase-mixed structures in Figure~\ref{fig:Rapo_Rperi} are clustered at low pericenters ($\lesssim 20$~kpc) and apocenters ($\lesssim 100$~kpc). They all have $f_{\rm bound,\star}\approx 0$, indicating their progenitors are entirely dissolved. As discussed in Section~\ref{subsubsec:streams}, these structures could potentially be resolved as streams at higher simulation resolution~\citep{Riley2025,Meziani}.  Therefore, the distribution of phase-mixed structures in the $R_{\rm apo}-R_{\rm peri}$ plane might suggest the potential for more coherent streams at smaller radii.  

For illustration, the black $\times$'s on the plot show current data from MW streams~\citep{Li_2022}. Ongoing surveys, such as the Rubin Observatory~\citep{Ivezic2019}, will probe further distances ($\sim300 \ \kpc$) for undiscovered stream populations~\citep{Bonaca2025}. To make robust predictions for these observations, one must properly account for the detectability of the streams by turning the simulation outputs into mock data catalogs~\citep{Shipp_2023, Kundu}. Additionally, given the substantial halo-to-halo variance on these predictions~\citep{2025ApJ...990..162D}, larger numbers of simulations are needed to properly quantify the spread in the theoretical predictions. 

\section{CONCLUSIONS}\label{sec:conclusion}

This work provides the first detailed study of how a dissipative DM model impacts the properties of stellar streams. As a concrete example, we focused on the example where the DM is comprised of CDM plus a subcomponent ($\sim 6\%$) of ADM. We ran a cosmological hydrodynamical zoom-in simulation of a MW-mass galaxy for this scenario (called \texttt{ADM-slow}), identifying stellar streams with $M_{\rm tot, \star}\gtrsim10^{5.5} \ \text{M}_\odot$ and intact satellites with $M_\star\gtrsim10^{4.5} \ \text{M}_\odot$.  The results were compared to those from a matching simulation consisting of only CDM (called \texttt{CDM}). 

The presence of an ADM subcomponent enhances the inner density of subhalos of $M_{200,\rm m}\gtrsim10^{8.25}\msun$. For the intact satellites and stream progenitors studied here, the total enclosed mass is enhanced within $\sim1$~kpc for \texttt{ADM-slow} compared to \texttt{CDM}. Additionally, the inner slope of the DM density profiles is also cuspier. The difference in inner subhalo density has important ramifications for its tidal disruption.  The mass-loss fraction of CDM for subhalos is comparable in both simulations and $\sim 69$\%, but the mass-loss fraction of ADM is far lower, $\sim 3$\%.  This is because the ADM clumps are more tightly bound in the central regions of the subhalo.  

The difference in subhalo densities in \texttt{CDM} and \texttt{ADM-slow} translates to differences in the stellar stream properties.  In the presence of an ADM subcomponent:
\begin{itemize}[itemindent=\dimexpr\labelwidth+\labelsep\relax,leftmargin=0pt]
    \item stream formation is delayed. This is likely because the progenitors are more centrally concentrated and thus more resistant to disruption.  
    \item stellar streams reach their peak $M_\star$ at later times after infall. This indicates that ADM prolongs their star formation histories, retaining gas for longer periods. 
    \item the chemical evolution of a stellar stream exhibits more frequent bursts of star formation over longer periods of time.  This is reflected in younger stars with [Fe/H] and [Mg/Fe] enhanced relative to the expectation in standard CDM.
    \item stream orbits are relatively unchanged, at least for the mass scales resolved here. However, more self-bound satellites with low $f_{\rm bound,\star}$ are identified at low $R_{\text{peri}}$, low $R_{\text{apo}}$ in \texttt{ADM-slow}. These objects may be the progenitors of unresolved streams, indicating more streams with surviving progenitors may be identified in the inner regions of the host in \texttt{ADM-slow} at higher resolutions.
\end{itemize}
These conclusions are based on a MW-mass halo generated for a specific initial condition (e.g., \texttt{m12i} in the FIRE suite). Future work should explore how the conclusions are affected by variations to the initial conditions, better quantifying the halo-to-halo variance.  

To resolve the streams in these simulations, we ran them at low baryon mass and high time resolution. We found nine stellar streams in both \texttt{ADM-slow} and \texttt{CDM}.  In addition, we identified a population of phase-mixed structures. A subset of these would likely be  reclassified as streams at higher resolution. This could potentially open up a population of streams within pericenters of $\sim 20$~kpc.  

This paper focused on shifts in stream properties relative to the case of standard CDM.  It intentionally did not address whether the shifts would be observable with current or future probes. While observational results for MW dwarfs and streams were included to put the results in a broader context, a detailed comparison to data would necessitate the creation of mock catalogs from the simulation outputs, which we save for future work. As underscored by \cite{Shipp_2023}, accounting for detectability in simulated stellar streams is critical for making robust comparisons to observations. This motivates future studies with mock Roman Space Telescope~\citep{Spergel2013} and Rubin Observatory~\citep{Ivezic2019} observations for comparisons with upcoming data.

While we focused on the chemical and orbital properties of the stellar streams, some of the findings could also have ramifications for other studies of small-scale structure. Due to the presence of ADM, the satellites in \texttt{ADM-slow} are more centrally concentrated than those in \texttt{CDM}. This also means that they survive to lower masses and lower pericenters. The presence of these dense subhalos can have ramifications for breaks in stellar streams. Such compact subhalos would be more likely to create gaps upon close encounters. Recent modeling based on observations of the GD-1 stream shows a preference for this enhanced compactness~\citep{Nibauer2025}, motivating models such as ADM. Additionally, subhalos with steep inner slopes can yield distinctive signatures in strong lensing studies, as discussed by~\cite{kollmann2025}. With inner slopes similar to those found in \texttt{ADM-slow}, subhalos with $M_{200,\rm m}\gtrsim 10^8 \ \msun$ would likely be detected by upcoming strong lensing surveys. Such data would allow for further constraints on this model.

The parameter space of minimal ADM is quite large, with five free parameters. This work focused on only one point in this space, motivated because it reduced the cooling rate of the ADM gas relative to other models studied in the literature~\citep{Roy:2024bcu}.  In more aggressively cooling  scenarios, the ADM can form rotating dark disks at the center of the MW-mass host~\citep{Roy:2023zar} and subhalos that are so concentrated they are clearly inconsistent with observations~\citep{Gemmell2024}. We found that the slowly cooling model is more consistent with the data on MW dwarfs~\citep{Kaplinghat2019}, although finer discrepancies remain.  For example, \texttt{ADM-slow} struggles to produce any cores in the satellites at all.  As more regions of ADM parameter space are probed in simulation---coupled with the generation of mock catalogs for better comparisons to data---it will enhance our ability to map out the viable regions of ADM parameter space.

In conclusion, this work demonstrated that a small subcomponent of strongly dissipative DM interactions can affect the properties of stellar streams at $z=0$.  These deviations are due to the centrally concentrated density of the subhalos in these models.  We expect that similar effects should occur in other models that affect subhalos in the same manner.  For example, weakly dissipative models, such as those studied by~\cite{Shen2021} and \cite{ONeil2023}, can lead to steep inner-density slopes ($\alpha \lesssim -3/2$). Additionally, SIDM with large scattering cross sections can lead to similarly cuspy profiles in the gravothermal collapse regime~\citep{Balberg2002}.  Conversely, opposite trends---such as lower resistance to tidal disruption---might arise in models that lead to cored profiles, such as  SIDM with low cross sections~\citep{spergel2000,Elbert2015}. Our results thus motivate a more comprehensive exploration of dark-sector models and their effects on stellar stream properties.  The improved understanding from such work will enable the community to harness stream data from upcoming surveys towards testing the particle nature of DM.

\section{Data Availability}

We provide a catalog of the simulated intact satellites, streams (and corresponding progenitors), and phase-mixed objects used in this study. For a given substructure, the data include all general properties analyzed in Sections~\ref{sec:props} and ~\ref{sec:streams}. The catalog is hosted at \href{https://doi.org/10.5281/zenodo.19135482}{https://doi.org/10.5281/zenodo.19135482}. Additional data will be shared on reasonable request to the corresponding author. Animations of our simulations are available \href{https://www.youtube.com/playlist?list=PL5IS42ggFs2iFXRQ8uSj1ZfRUGwdibxMM}{here}.


\section{Acknowledgments}
The authors would like to acknowledge helpful conversations and feedback from Matt Coleman, Akaxia Cruz, Caleb Gemmell, Jiaxuan Li, Ethan Lilie, Jonah Rose, and Andrew Wetzel. 

LMG, SO, ML, and SR are supported by the National Science Foundation~(NSF), under Award Number AST~2307789. ML is also supported by the Simons Investigator in Physics Award. AA is supported by the Gordon and Betty Moore Foundation. LN is supported by the Sloan Fellowship, the NSF CAREER award 2337864, and the NSF award 2307788. XS is supported by the NASA theory grant JWST-AR-04814. This research was also supported in part by grant NSF PHY-2309135 to the Kavli Institute for Theoretical Physics (KITP). The work reported on in this paper was also partly performed using the Princeton Research Computing resources at Princeton University which is consortium of groups led by the Princeton Institute for Computational Science and Engineering~(PICSciE) and Office of Information Technology's Research Computing. The authors also acknowledge the Texas Advanced Computing Center~(TACC) at The University of Texas at Austin for providing computational resources that have contributed to the research results reported within this paper.

%

\vspace{5mm}


\software{\textsc{Jupyter~\citep{Kluyver2016jupyter}, \textsc{matplotlib}~\citep{Hunter:2007}, \textsc{NumPy}~\citep{harris2020array}, \textsc{SciPy}~\citep{Virtanen2020}, \textsc{pyFalcON}~\citep{Dehnen2000}, \textsc{SWIFTsimIO}~\citep{Borrow2020,Borrow2021}, \textsc{unyt}~\citep{Goldbaum2018}, \textsc{GizmoAnalysis}~\citep{Wetzel2020}, \textsc{HaloAnalysis}~\citep{Wetzel2020halo}}
          }



\bibliography{main}{}
\bibliographystyle{aasjournal}

\appendix

\setcounter{equation}{0}
\setcounter{figure}{0}
\setcounter{table}{0}
\renewcommand{\theequation}{A\arabic{equation}}
\renewcommand{\thefigure}{A\arabic{figure}}
\renewcommand{\thetable}{A\arabic{table}}
\renewcommand{\theHequation}{A\arabic{equation}}
\renewcommand{\theHfigure}{A\arabic{figure}}
\renewcommand{\theHtable}{A\arabic{table}}

\section{SUPPLEMENTARY FIGURES}\label{sec:main}

Here, we present supplementary figures that provide further context to results from the main text. 

\begin{figure*}[h]
\centering
\includegraphics[width=0.6\textwidth]{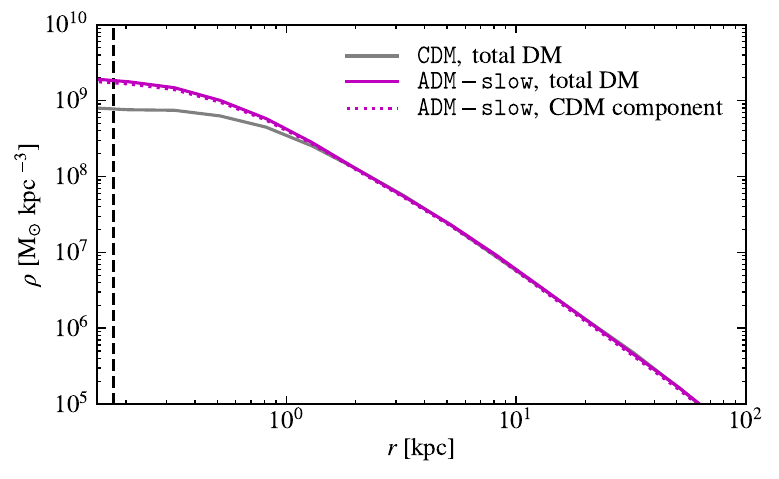}
\caption{Dark matter~(DM) density profiles for the Milky Way~(MW)-mass hosts in \texttt{CDM}~(gray) and \texttt{ADM-slow}~(magenta) at $z=0$. The solid lines represent the profiles for the total DM density.  The dotted magenta line corresponds to the profile of the CDM component in \texttt{ADM-slow}. The vertical dashed line shows the convergence radius $r^{\text{conv}}_{\text{dm}}$ for \texttt{CDM}, which is slightly higher than that for \texttt{ADM-slow}. Both profiles largely agree for $r \gtrsim 1 \ \text{kpc}$, but \texttt{ADM-slow} shows an enhanced inner density. This enhancement comes from the CDM responding to the deeper potential well due to the presence of the ADM.}
\label{fig:main_densities}
\end{figure*}

\begin{figure*}[h]
\centering
\includegraphics[width=\textwidth]{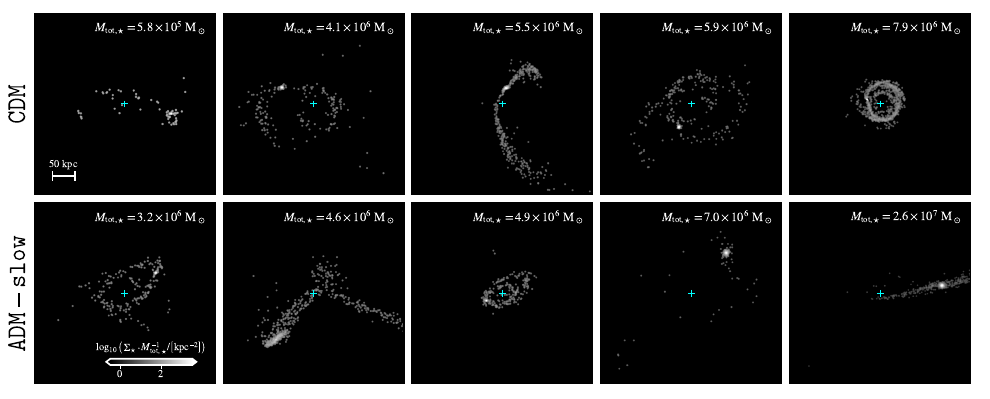}
\caption{Density projections for the stars in the remaining stellar streams at $z=0$ not shown in Figure~\ref{fig:streams}. The top row corresponds to streams in \texttt{CDM} and the bottom to \texttt{ADM-slow}. Each projection is shown from the same angle, centered at the main halo center, indicated as a cyan cross. The present-day stellar mass of each stream is shown on the corresponding plot, with mass increasing from left to right. Animations showing the formation of these substructures are available \href{https://www.youtube.com/playlist?list=PL5IS42ggFs2iFXRQ8uSj1ZfRUGwdibxMM}{here}.}
\label{fig:remaining_streams}
\end{figure*}

\begin{figure*}[t!]
\centering
\includegraphics[width=\textwidth]{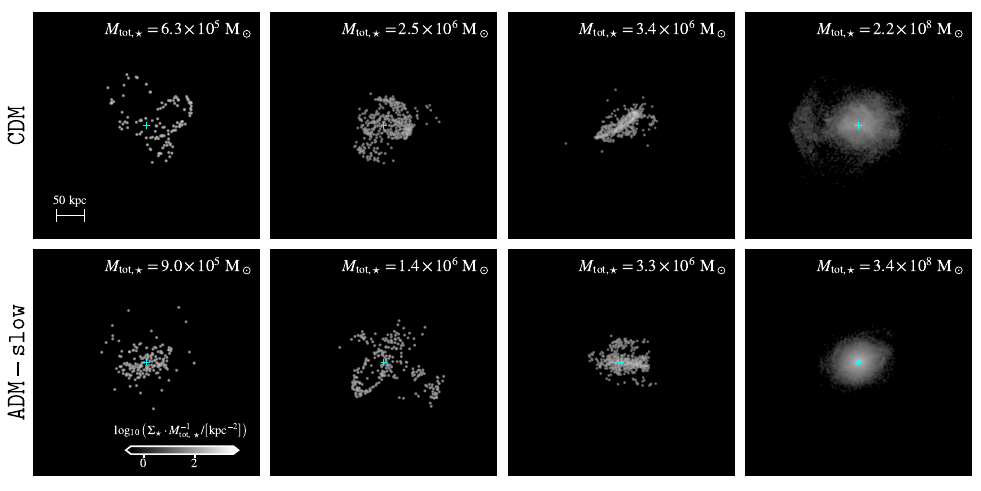}
\caption{Examples of density projections for the stars in phase-mixed structures at $z=0$. The top row corresponds to structures in \texttt{CDM} and the bottom to \texttt{ADM-slow}. Each projection is shown from the same angle, centered on the main halo, indicated by the cyan cross. The present-day stellar mass of each structure is shown on the corresponding plot, with mass increasing from left to right. All structures have median local velocity dispersions above the relation defined in Equation~\ref{eq:criterion}, though some lie very close to the line and are ambiguous. Animations showing the formation of these substructures are available \href{https://www.youtube.com/playlist?list=PL5IS42ggFs2iFXRQ8uSj1ZfRUGwdibxMM}{here}.}
\label{fig:pm}
\end{figure*}

\begin{figure*}[t!]
\centering
\includegraphics[width=\textwidth]{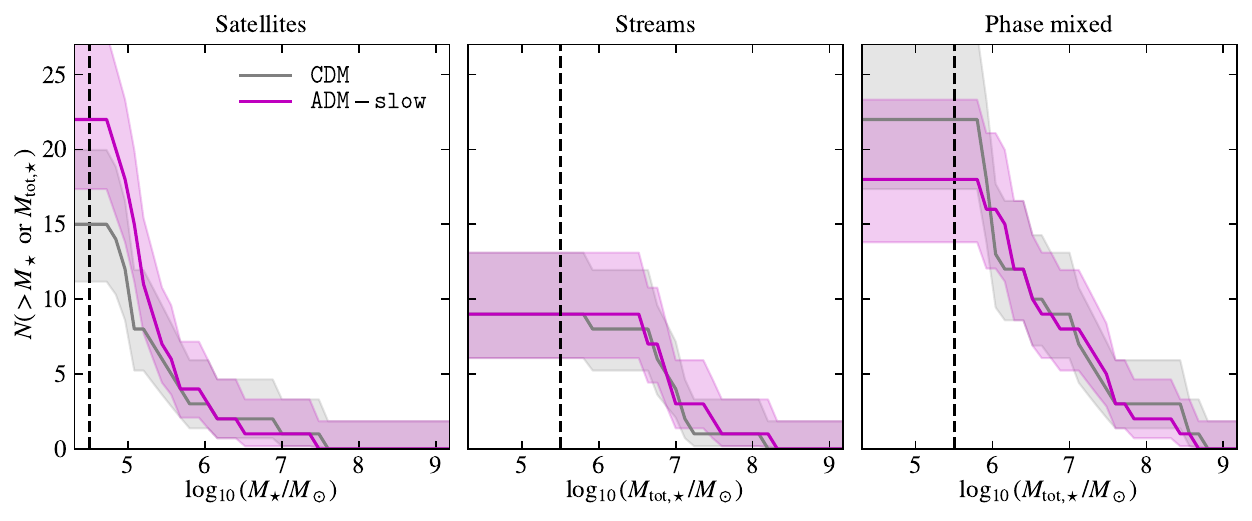}
\caption{Mass functions for the intact satellites~(left panel), streams~(middle panel), and phase-mixed structures~(right panel) in  \texttt{CDM}~(gray) and \texttt{ADM-slow}~(magenta) at $z=0$. Stellar masses are based on procedures described in Section~\ref{subsec:id_streams}. Shaded bands indicate the $1\sigma$ Poisson error for each bin. The vertical dashed lines indicate the effective stellar mass cut for each population based on the selection criteria from Section~\ref{subsec:id_streams}.  The satellite mass function is enhanced in \texttt{ADM-slow} below $M_\star \lesssim 10^{5.5} \ \msun$ relative to that in \texttt{CDM}.  For streams, the mass functions are consistent between the two simulations, within Poisson scatter.  The mass distribution of phase-mixed structures is also indistinguishable between the two simulations, although there is a hint of an enhancement in \texttt{ADM-slow} below $M_{\rm tot, \star} \lesssim 10^6 \ \msun$.}
\label{fig:mass_functions}
\end{figure*}

\begin{figure*}[t!]
\centering
\includegraphics[width=\textwidth]{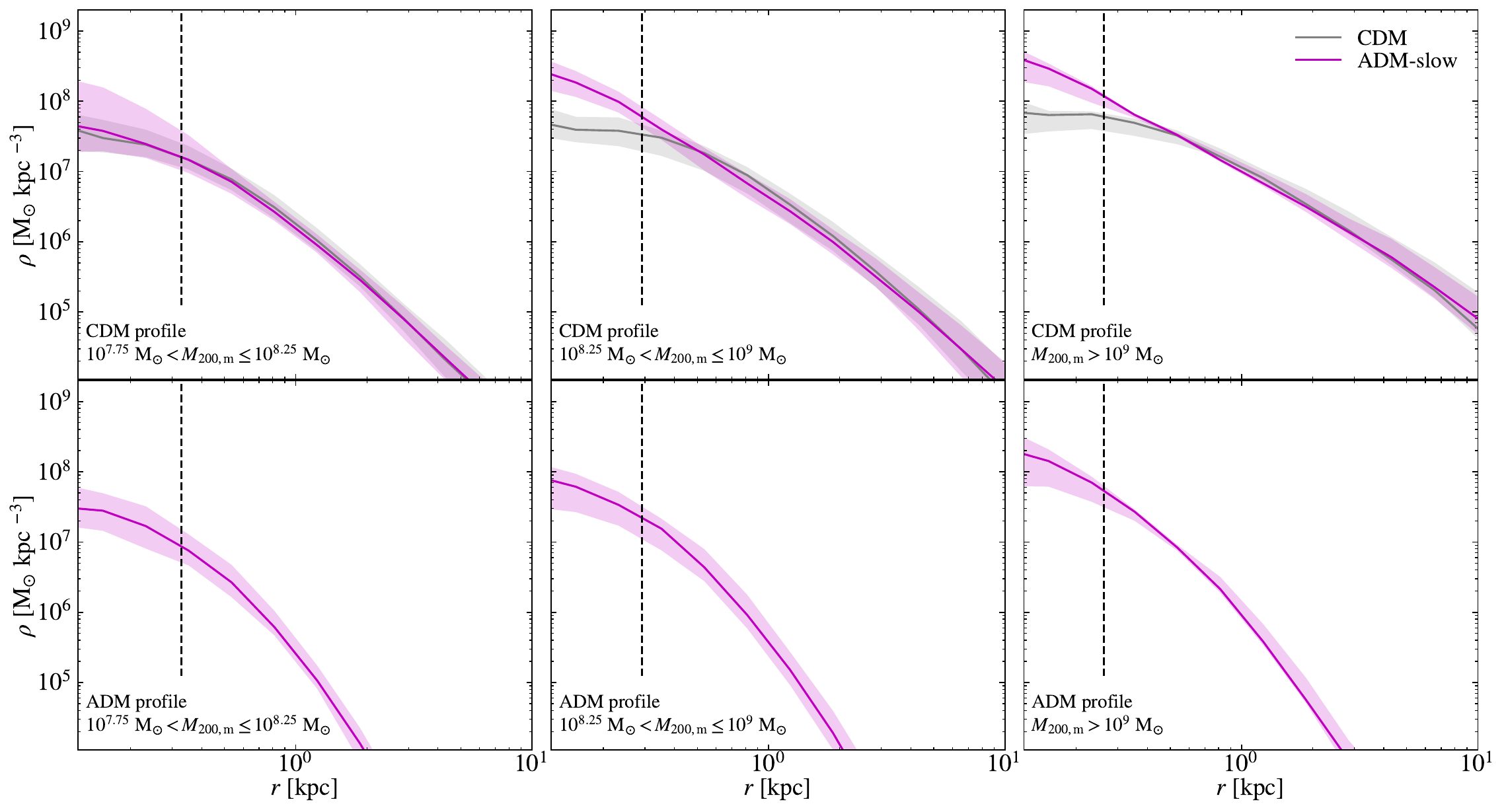}
\caption{\textbf{Top:} Median CDM density profiles for all subhalos, per radial bin, at $z=0$. Shaded bands indicate the $68\%$ containment regions. Density profiles are calculated for subhalos in three different mass bins. From left to right: $10^{7.75} \ \text{M}_\odot < M_{\text{200,m}}\leq 10^{8.25} \ \text{M}_\odot$, $10^{8.25} \ \text{M}_\odot < M_{\text{200,m}}\leq 10^{9} \ \text{M}_\odot$, and $M_{\text{200,m}}> 10^{9} \ \text{M}_\odot$. The vertical dashed lines indicate the highest $r_{\text{dm}}^{\text{conv}}$ for the corresponding mass bin. \texttt{ADM-slow} subhalos show a slight enhancement of the CDM inner slope, likely due to adiabatic contraction from ADM gas. \textbf{Bottom:} Median ADM clump density profiles for ADM-containing subhalos, per radial bin, at $z=0$. ADM clumps consistently contribute less to the overall density than CDM outside of $\gtrsim 1 \ \kpc$.}
\label{fig:sub_densities}
\end{figure*}

\begin{figure*}[t!]
\centering
\includegraphics[width=0.81\textwidth]{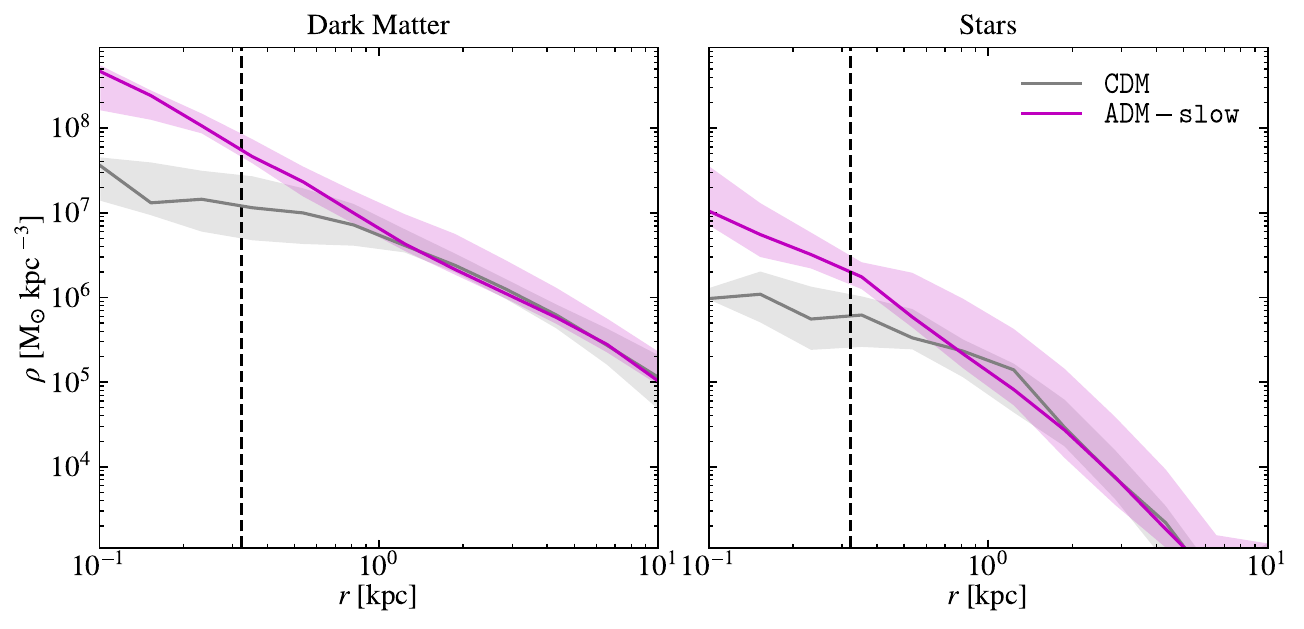}
\caption{\textbf{Left:} Median DM (CDM+ADM clumps for \texttt{ADM-slow}) density profiles for all stream progenitors, per radial bin, at their time of peak stellar mass ($\tau_{\rm peak}$). Shaded bands indicate the $68\%$ containment regions. The vertical dashed lines indicate the highest $r_{\text{dm}}^{\text{conv}}$ among all progenitors. \texttt{ADM-slow} progenitors show enhanced DM inner densities, both from direct contributions from ADM and from enhancements to CDM. \textbf{Right:} Median stellar density profiles for all progenitors, per radial bin, at $\tau_{\rm peak}$. The progenitors in \texttt{ADM-slow} show enhanced inner densities in the stellar population as well.}
\label{fig:prog_densities}
\end{figure*}

\begin{figure*}[t!]
\centering
\includegraphics[width=0.5\columnwidth]{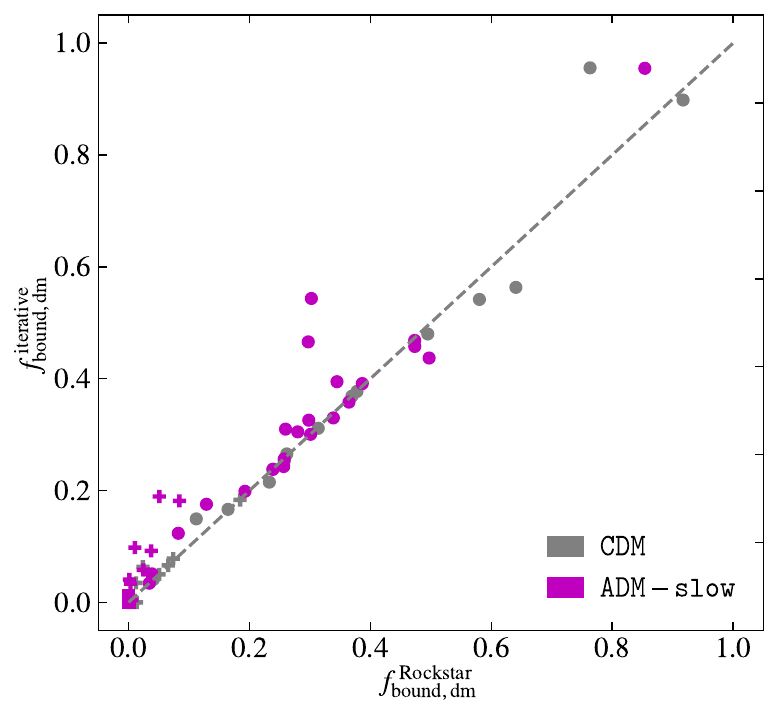}\includegraphics[width=0.5\columnwidth]{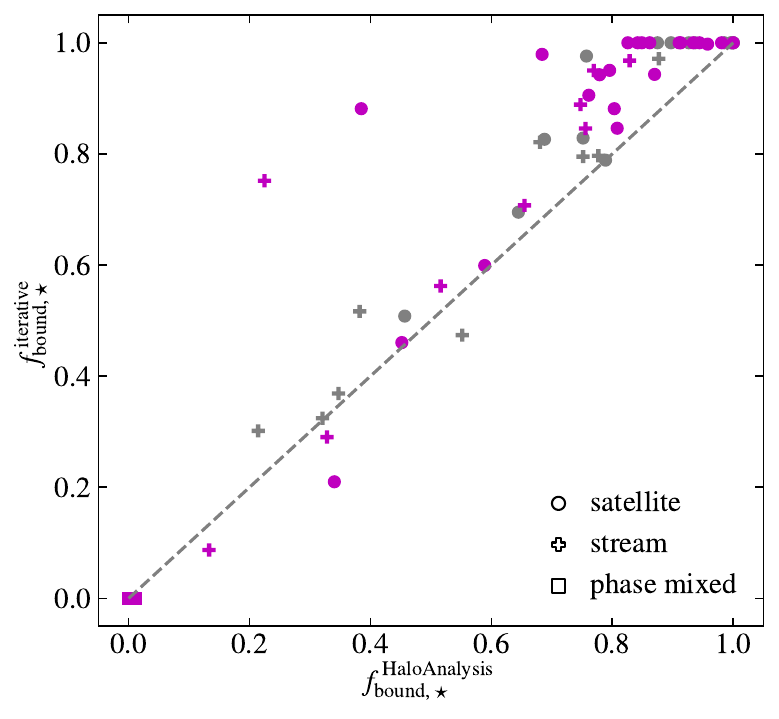}
\caption{\textbf{Left:} Comparison between DM bound mass fraction $f_{\rm bound, dm}=M_{\rm bound, dm}(z=0)/M_{\rm peak}$ for all substructures as calculated through iterative unbinding and \textsc{Rockstar}. The former procedure is described in Section~\ref{subsec:stripping}, while the latter determines bound particles through single pass unbinding, assuming a spherical potential. Here, $M_{\rm bound, dm}(z=0)$ is the DM mass bound to a given halo at $z=0$ and $M_{\rm peak}$ represents the peak $M_{200,\rm m}$ for the halo. Both methods are generally consistent, but iterative unbinding is able to recover more particles for a few systems, especially those with ADM clumps. \textbf{Right:} Comparison between stellar bound mass fraction $f_{\rm bound, \star}=M_{\star}(z=0)/M_{\rm tot,\star}$ for all substructures as calculated through iterative unbinding and \textsc{HaloAnalysis}. Here, $M_{\star}(z=0)$ is the stellar mass bound to a given halo at $z=0$ and $M_{\rm tot,\star}$ represents the total present-day stellar mass of all stars that have been assigned to a halo throughout its evolution. Both methods are often consistent, but iterative unbinding is able to recover significantly more star particles for less-disrupted systems.
\label{fig:fbound}}
\end{figure*}

\begin{figure*}[t!]
\centering
\includegraphics[width=0.5\columnwidth]{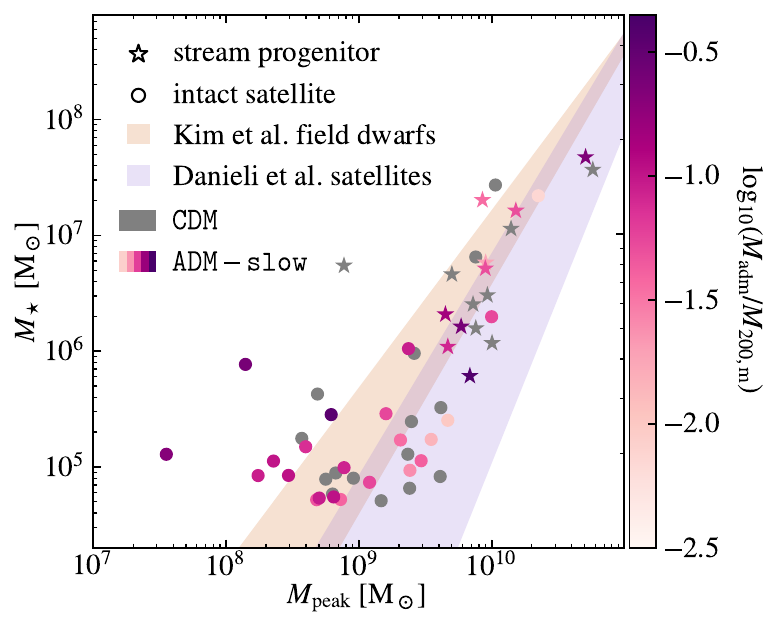}
\caption{Stellar-to-halo mass relation ($M_\star$ versus peak $M_{\text{200,m}}$) for the satellites~(circle markers) and stream progenitors~(star markers). $M_\star$ values represent the bound stellar mass at $z=0$ based on \textsc{HaloAnalysis}. The \texttt{ADM-slow} markers are shaded based on the ADM clump mass fractions of the corresponding subhalos.  The \texttt{CDM} markers are shaded gray. The orange band represents the best-fit line and scatter for field dwarfs found by \citet{Kim2024}. The purple band represents the extrapolated trend for dwarf satellites found by \citet{Danieli2023}. Satellites in $\texttt{ADM-slow}$ can survive to lower $M_{\rm peak}$ than those in \texttt{CDM}, likely because they contain a dense inner core of ADM clumps that is more resistant to tidal disruption. 
\label{fig:SMHMR}}
\end{figure*}

\begin{figure*}[t!]
\centering
\includegraphics[width=\textwidth]{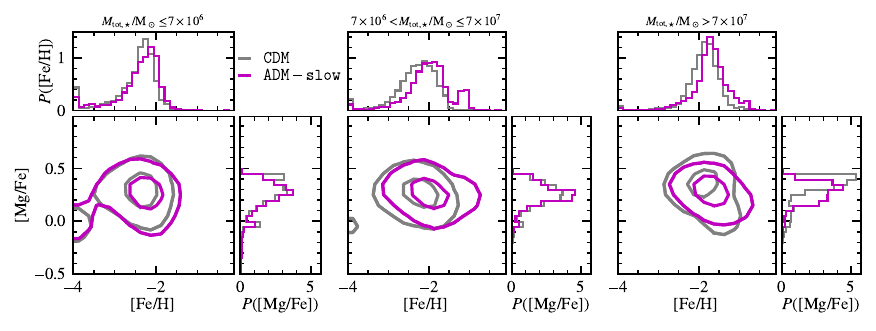}
\caption{[Mg/Fe]$-$[Fe/H] tracks for all stars associated with present-day streams at $z=0$. The 2D contours represent the 1 and $2\sigma$ containment regions, smoothed with a Gaussian filter.  Results for \texttt{CDM} are shown in gray and for \texttt{ADM-slow} in magenta. Each set of histograms corresponds to a different mass bin: $M_{\rm tot, \star} \leq 7\times 10^6\,\text{M}_\odot$~(left), $7\times10^6\,\text{M}_\odot < M_{\rm tot, \star} \leq 7\times 10^7\,\text{M}_\odot$~(middle), and $M_{\rm tot, \star} > 7\times 10^7\,\text{M}_\odot$~(right). The right-most column represents the most-massive stream from each simulation, which has significantly more stars than all the others. The 1D projections of the [Fe/H] and [Mg/Fe] distributions are provided along the top and right axes, respectively, of each panel. }
\label{fig:MgFe}
\end{figure*}

\begin{figure*}[t!]
\centering
\includegraphics[width=0.475\textwidth]{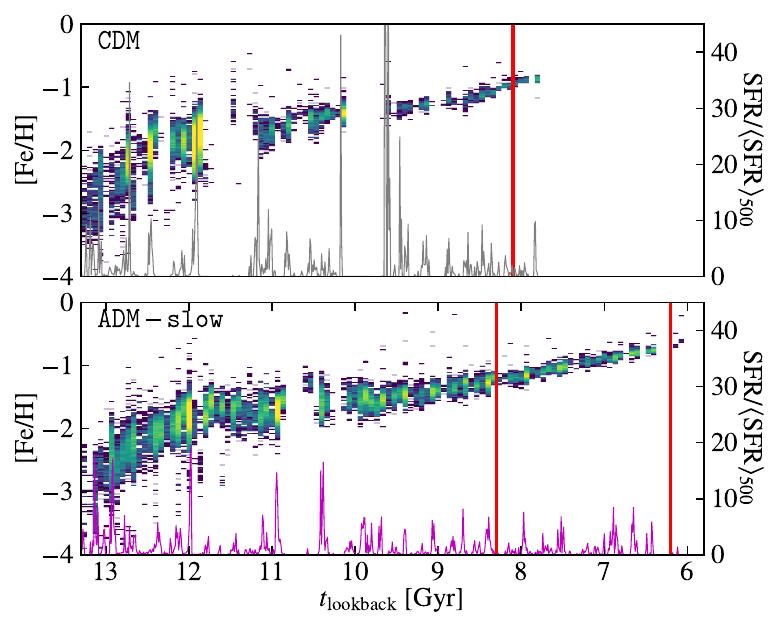}
\includegraphics[width=0.47\textwidth]{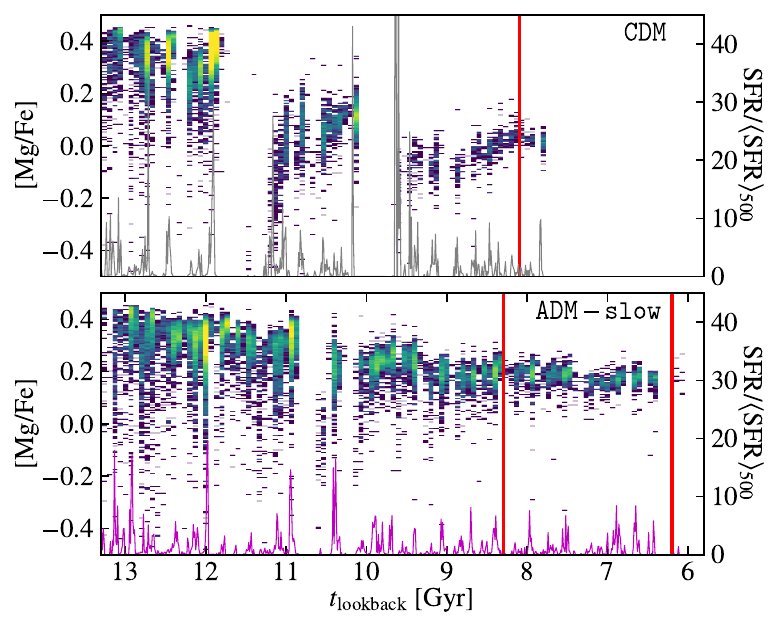}
\caption{Chemical evolution and normalized star formation rate~(SFR) for the most massive stream found in each simulation ($M_{\rm tot, \star} = 1.4\times 10^8 \ \text{M}_\odot$ in \texttt{ADM-slow} and $M_{\rm tot, \star} = 1.1\times 10^8 \ \text{M}_\odot$ in \texttt{CDM}). The left plots show the 2D projections of the [Fe/H] of the stars over their formation history and the right plots show the corresponding projections of [Mg/Fe]. The top plots show the results for \texttt{CDM}, with the star formation rate~(SFR) in gray. The bottom plots show the results for \texttt{ADM-slow}, with the SFR in magenta. The SFR is normalized by the average SFR over a retrospective window of $500$~Myr. The red vertical lines correspond to times of closest pericentric approach. The stream in \texttt{ADM-slow} has a longer star formation history, continuing past the first pericentric approach. In general, the youngest stars have the highest [Fe/H] and lowest [Mg/Fe]. The stream in \texttt{CDM} shows long periods of quiescence, which lead to sudden dips in [Mg/Fe] that are then partially recovered as new Type~II supernovae are introduced. These dips are not visibly present in the [Fe/H] distribution, due to Fe being formed over long timescales. With shorter periods of quiescence, the [Mg/Fe] in \texttt{ADM-slow} is replenished more continuously.}
\label{fig:SFR}
\end{figure*}

\end{document}